\documentclass[12pt,preprint]{aastex}

\def\p/{\mbox{$^1$}}
\def\pp/{\mbox{$^2$}}
\def\ppp/{\mbox{$^3$}}
\def\pppp/{\mbox{$^4$}}
\def\m/{\mbox{$^{-1}$}}
\def\mm/{\mbox{$^{-2}$}}
\def\mmm/{\mbox{$^{-3}$}}
\def\mmmm/{\mbox{$^{-4}$}}
\def\Ms/{\mbox{M$_\odot$}}

\shorttitle{IMF biases created by blending}
\shortauthors{Ma\'{\i}z Apell\'aniz}

\slugcomment{Accepted for publication in the Astrophysical Journal}

\begin{document}

\title{Biases on initial mass function determinations. II. Real multiple systems and chance superpositions\altaffilmark{1}}

\author{J. Ma\'{\i}z Apell\'aniz\altaffilmark{2,3}}
\affil{Instituto de Astrof\'{\i}sica de Andaluc\'{\i}a-CSIC, Camino bajo de Hu\'etor 50, 18008 Granada, Spain}


\altaffiltext{1}{This article is partially based on observations made with the NASA/ESA \facility{Hubble Space Telescope} (HST), 
some of them associated with GO program 10602 and the rest gathered from the archive, obtained at the Space Telescope Science 
Institute, which is operated by the Association of Universities for Research in Astronomy, Inc., under NASA contract NAS 5-26555.}
\altaffiltext{2}{e-mail contact: {\tt jmaiz@iaa.es}.}
\altaffiltext{3}{Ram\'on y Cajal fellow.}

\begin{abstract}
When calculating stellar initial mass functions (IMFs) for young clusters, one has to take into account that
most massive stars are born in multiple systems and that most IMFs are 
derived from data that cannot resolve such systems. It is also common to 
measure IMFs for clusters that are located at distances where multiple chance superpositions between members are expected to happen. In this 
article I model the consequences of both of those phenomena, real multiple systems and chance superpositions, on the observed color-magnitude diagrams and the IMFs 
derived from them. Using numerical experiments I quantify their influence on the IMF slope for massive stars and on the generation of systems with
apparent masses above the stellar upper mass limit. The results in this paper can be used to correct for the biases induced by real and chance-alignment
multiple systems when the effects are small and to identify when they are so large that most information about the IMF in the observed color-magnitude diagram
is lost. Real multiple systems affect the observed or apparent massive-star MF slope little but can create a significant population of apparently ultramassive stars.
Chance superpositions produce only small biases when the number of superimposed stars is low but, once a certain number threshold is reached, they can affect
both the observed slope and the apparent stellar upper mass limit.
In the second part of the paper, I apply the experiments to two well known massive young clusters in the Local Group, NGC 3603 and R136.
In both cases I show that the observed population of stars with masses above 120 \Ms/ can be explained by the effects of unresolved objects, mostly
real multiple systems for NGC 3603 and a combination of real and chance-alignment multiple systems for R136. Therefore, the case for the reality of a
stellar upper mass limit at solar or near-solar metallicities is strengthened, with a possible value even lower than 150 \Ms/. 
An IMF slope somewhat flatter than Salpeter or Kroupa with $\gamma$ 
between $-1.6$ and $-2.0$ is derived for the central region of NGC 3603, with a significant contribution to the uncertainty arising from the imprecise 
knowledge of the distance to the cluster. The IMF at the very center of R136 cannot be measured with the currently available data but the situation could 
change with new Hubble Space Telescope (HST) observations. 
\end{abstract}


\section{Introduction}

	This paper is the second one of a series where we explore the effects of different biases on the determination of the stellar initial 
mass function (IMF). In paper I \citep{MaizUbed05} we analyzed the numerical biases induced by 
using bins of equal width when fitting power-laws to binned data (an effect that is more general than its application to the calculation of
mass functions). Those biases can be eliminated in several ways, of which a simple one is by grouping the data in
equal-number bins (as opposed to equal-width bins). In this second paper I explore the effect of unresolved multiple systems, either physical or
chance alignments, especially for the high-mass end of the IMF. In a future paper we will analyze the effect of random uncertainties in 
the mass determinations on the computed initial mass function\footnote{Most of the analysis in these papers is also applicable to the
calculation of the present-day mass function (PDMF).}.

\subsection{Binaries and the IMF}

	Why worry about binaries and higher-order multiple systems
when calculating the IMF of a young stellar cluster such as NGC 3603 or R136? The quick answer is because 
one expects most of the observed point sources in clusters beyond a few kpc to be multiple. Therefore, deriving their masses by assuming they are single 
stars should introduce biases in the determination of the stellar mass functions of the clusters. 

	\citet{Masoetal98} studied nearby O stars from a
sample that included field objects and stars in low- and intermediate-mass clusters and associations. 
They analyzed the observed multiple fractions as a function of environment, including both spectroscopic and visual/speckle binaries and discovered 
that at least 72\% of the O stars in clusters and associations had
detected companions. Furthermore, the period distribution had a clear bimodal structure, with one peak around 10 days and another one around
$10^5$ years. They argued that such an effect was quite likely to be instrumental because binaries with periods $P$ near 1-1000 years are hard to detect
and because the period distribution for B stars is quite flat\footnote{Such a distribution, expressed in terms of separation $r$ instead of $P$, is close to
$f(\log r) \propto (\log r)^0$ and is known as \"Opik's law, after \citet{Opik24}.}. Therefore, \citet{Masoetal98} predicted that when better spatial 
resolution were achieved and more intensive radial-velocity variation surveys were undertaken in the future, the gap between the two peaks might be 
filled and the true multiple fraction for massive stars in clusters might be found to be close to unity. 
Another interesting result of that paper was that the mass-ratio distribution for O binaries was
flat or nearly so, i.e. massive stars tend to have companions that are more massive on average than what would be expected from a random sampling
of the IMF. The shape of the secondary mass distribution function is relevant because different slopes should affect the observed or apparent mass function
derived from unresolved data in different ways. \citet{SagaRich91} calculated that the apparent MF slope for intermediate-mass stars in the random pairing case
becomes significantly flatter when adding unresolved binaries but the effect is likely to be different in the flat mass-ratio case.

	In the ten years after \citet{Masoetal98} other works have validated their conclusions. \citet{GarcMerm01} measured that 11 
out of 14 O-systems in the cluster NGC 6231 are spectroscopic binaries. This means that at least 79\% of its O/WR systems
are multiple and at least 88\% of their O/WR (and some early B) stars are in multiple systems. Their numbers are also similar for
early-B stars. Also, new studies are indeed filling the Mason period gap: \citet{Gameetal07} reported 11 new massive spectroscopic binaries and 
\citet{Maizetal08b} cite two new visual systems in the solar neighborhood (see also \citealt{Nelaetal04}). With respect to the secondary mass 
distribution function, \citet{Kouwetal05} studied 
intermediate-mass stars in the Sco OB2 association and found that they could exclude random pairing in binaries, since the more massive stars clearly 
tend to associate with other massive objects. Another recent study of Cyg OB2 by \citet{KobuFrye07} also finds a large ($>$80\%) binary fraction for
massive stars. Those authors favor a secondary mass distribution function that is either (a) nearly flat or (b) a combination of a 60\% Kroupa-like 
distribution and a 40\% ``twin'' ($M_2 > 0.95 M_1$) component.

	But what about more massive clusters? Is it possible that the multiple fraction there is significantly lower? Studying
distant objects is more complicated but a recent survey by \citet{Evanetal06} found that NGC 346 (in the SMC), NGC 2004, and N11 (in the LMC)
have a minimum spectroscopic binary fraction for O- and early B-type stars of 23 to 36\%. Considering that their program did not allow them to detect 
binaries with periods of more than several tens of days and that at the distance of the Magellanic Clouds it is very hard to resolve visual binaries, 
the real binary fraction should be considerably higher. Therefore, it is safe to conclude that the effect of unresolved massive binaries in the 
observed IMF is important and needs to be analyzed. Such an analysis is the first goal of this paper.

\subsection{The stellar upper mass limit}

	From the point of view of the origin of the IMF, it has become apparent in the last decade that massive stars must form in a different way 
than low-mass objects because of the short time scales involved and the limiting effect of radiation pressure on the mass accretion rate
\citep{BallZinn05,ZinnYork07}. There are currently two viable alternatives to form a massive star: by accretion through a disk, either in a 
relatively isolated environment or in a more crowded one where competition for mass among different protostars may be important, and by stellar 
collisions followed by mergers. The second alternative can only be relevant when the stellar densities are very high, such as at the core of a
dense and massive cluster, but cannot be significant in relatively low-density environments such as OB associations. 

	If stellar mergers are indeed important at the core of dense clusters, there should be large consequences on their stellar IMFs.
\citet{Portetal99} predicted that runaway collisions in such cores will lead to stars with masses above 100 \Ms/ in a time scale of one or a few 
million years, which is comparable to the expected life times of those stars. The process may keep going on to the point of producing objects 
above 1000~\Ms/ that would end up as intermediate-mass black holes \citep{Portetal04a}. Therefore, massive dense clusters should 
show an stellar upper mass limit $m_{\rm max}$ significantly higher than those of lower density or mass. Can such stellar behemoths exist? Recently, 
\citet{Belketal07} and \citet{Yungetal08} have attempted modeling of solar-metallicity stars up to 1000~\Ms/, which would be formed by 
mergers. However, such modeling encounters problems with the determination of the Eddington limit, its associated instabilities and
interaction with rotation, and with the large extrapolation required for mass-loss rates (which are already uncertain at the present time even
for run-of-the-mill main-sequence O stars). Hence, theory has not given its final word on the
possible existence of such ultramassive stars.

	What do observations tell us about those hypothetical ultramassive stars? Several papers on R136 and other Local Group massive young clusters 
\citep{WeidKrou04,Fige05,OeyClar05,Koen06} indicate that $m_{\rm max} \approx$ 120-200 \Ms/ when evolutionary tracks are used to
obtain stellar masses\footnote{Those values are called evolutionary masses and refer to the initial mass of the star.}. Some luminous blue variables 
(LBVs) such as $\eta$~Car could have slightly higher evolutionary masses but LBVs are notoriously difficult to observe in detail due to the 
existence of circumstellar material. With respect to the more accurate masses measured 
from the orbital motion of the stars (keplerian masses), the highest published values correspond to WR 20a, a binary with 83.0~$\pm$~5.0~\Ms/ and 
82.0~$\pm$~5.0~\Ms/ (\citealt{Bonaetal04}, see also \citealt{Rauwetal04}). Therefore, on a first impression it appears that observations
do not favor the existence of ultramassive stars at near-solar metallicity.

	Unresolved multiple systems can also play a role in the identification of ultramassive stars by making e.g. a binary made
out of two stars close to the stellar upper mass limit appear as a single object with an evolutionary mass well above that limit. For example, 
\citet{Maizetal07} found that Pismis 24-1, formerly an ultramassive star candidate, is in reality composed of three stars with masses below 120 \Ms/.
It is also possible that the observed $m_{\rm max}$ measured from evolutionary masses is 
affected not only by real (bound) multiple systems but also by chance alignments in the crowded environments where some of these
objects are found. Analyzing the possibility that some of the objects observed close to $m_{\rm max}$ may be blended sources composed of
stars of significantly lower mass is the second goal of this paper.

\section{Experiments}

	In order to test the effect of unresolved binaries and blending on the massive-star IMF slope and the stellar upper mass limit I have designed 
three numerical experiments. The first one analyzes the effect of real binaries, the second one that of chance superpositions, and the third one
combines the two effects. The purpose of the experiments is to analyze the effects on a simple test IMF, provide a way to correct them, and describe
the tools needed to extend the technique to other circumstances.

	I design the simple case by assuming a well-sampled single stellar population with an age of 1 million years and solar metallicity. The
choices for the age and metallicity are given by our desire to test our results with an analysis of NGC 3603 and R136 (note that the latter is currently
thought to be of somewhat lower metallicity and may be slightly older) and by the simplicity of not having objects clearly
evolved off the main sequence or even collapsed. Since we are only interested in the behavior of the top part of the IMF, for simplicity I assume 
that all the stars have already reached the main sequence, even though such an assumption should not valid for the low-mass stars. The $T_{\rm eff}$, 
$\log g$, and $\log L$ information was obtained from the Geneva evolutionary tracks and isochrones 
database by \citet{LejeScha01} selecting the case with standard mass-loss rates and no rotation\footnote{In the last years much work has been done 
on new evolutionary tracks and isochrones with rotation and possibly-reduced mass-loss rates due to clumping corrections. The present work can 
be easily adapted to those other conditions but the effects on our results are expected to be
relatively minor.}. For masses below 0.8 \Ms/ I use the $M_V$ calibration of \citet{Krouetal93} to extend the isochrone until 0.1 \Ms/. For the
value of $m_{\rm max}$, I select the highest value available in the Geneva tracks, 120 \Ms/, which is on the low end of the possible range (for $m_{\rm max}$)
for solar and near-solar metallicities. The 1-million-year isochrone was inserted into 
the new evolutionary synthesis module of CHORIZOS \citep{Maiz04c} to calculate the Johnson $U$ and $V$ absolute magnitudes\footnote{For the sake of notation
simplicity and since at this point I am not discussing distance effects, I write $M_U$ and $M_V$ simply as $U$ and $V$, respectively.}
as a function of mass using different synthetic spectral energy distributions (SEDs) as a function of $T_{\rm eff}$: TLUSTY \citep{LanzHube03,LanzHube07}, 
Kurucz \citep{CastKuru03}, and Lejeune \citep{Lejeetal98} for the high, intermediate, and low temperature ranges, respectively. The Johnson zero
points derived by \citet{Maiz06a} and modified by \citet{Maiz07a} were used. The final results are two tabulated, finely-gridded 
functions that give $V$ and $U-V$ as a
function of mass $m$ between 0.1 and 120 \Ms/. The two functions were then extrapolated in $\log m$ up to $m = 1000$ \Ms/ ($\log m = 3$). Note
that $V(m)$ is a strict monotonically decreasing function over the full range of $m=0.1-1000$ \Ms/ (this is true for the chosen isochrone but it
is not true in general). Therefore, it is possible to invert it to produce $m(V)$, which gives the true mass\footnote{Throughout
this paper $m$ is considered to be the initial stellar mass, not the mass when the star is 1 million years old.} that corresponds to a 
star of absolute magnitude $V$.

	Once $V(m)$ and $U(m)-V(m)$ as a function of mass have been generated, I calculate the corresponding bidimensional single-star 
color-magnitude density function $g_{1{\rm s}}(U-V,V)$ by applying one of two cases: [a] a Kroupa IMF, which can be described as a continuous function 
comprised of two power law segments $dn/dm=f(m)\propto m^\gamma$, with $\gamma = -1.3$ for $m=0.1-0.5$ \Ms/ and $\gamma = -2.3$ for 
$m=0.5-120$ \Ms/; [b] a flatter, top-heavy IMF, similar to Kroupa but with $\gamma = -1.3$ for $m=0.1-0.5$ \Ms/ and  $\gamma = -2.0$ for $m=0.5-120$ \Ms/. 
$g_{1{\rm s}}(U-V,V)$ can be collapsed into a 
one-dimensional magnitude function\footnote{In this paper I use magnitude functions instead of luminosity functions; the conversion between one and the
other is straightforward.} $G_{1{\rm s}}(V)$ without losing any information because for our isochrone
$V(m)$ is a strictly monotonic function. Since both $g_{1{\rm s}}(U-V,V)$ and $G_{1{\rm s}}(V)$ are functions defined in a fine
(2-D and 1-D, respectively) grid, special care has to be taken to avoid introducing numerical biases and noise due to sampling and interpolation.

	Finally, in order to reproduce the blending problems associated with the strategy commonly used to derive evolutionary masses from luminosity 
functions, I combine individual stars into blended objects according to different rules. This is done by adding their $U$- and
$V$-band luminosities and computing the resulting $V$ and $U-V$ values. From there, I generate: [a] a new bidimensional
color-magnitude density function $g_{x}(U-V,V)$ from $g_{1{\rm s}}(U-V,V)$, where $x$ specifies the combination rule, and [b] the apparent mass 
function $f_{a,x}(m_a)$ calculated by collapsing $g_{x}(U-V,V)$ into $G_{x}(V)$ and then converting the observed absolute magnitudes into apparent masses. 
The last step is done by applying $m(V)$, which is a relationship derived for individual stars that is not correct for blended objects. Therefore,
the derived apparent masses $m_a$ are not the real ones. In other words, I am deriving an apparent mass function (AMF)
by [a] assuming that all the observed objects are single stars (even though they are not) and [b] using only the absolute magnitudes (but not 
their colors) to obtain the masses. The first assumption is applied in most IMF studies. The second assumption should be valid as long as all the stars are 
at the same distance, extinction is well known, contamination by field objects or stars of different ages is negligible, and $m(V)$ is a strictly monotonic
function. Note, however, that for ages older than $\approx 2.5$ million years (when $m(V)$ is no longer monotonic) the same type of analysis can still be done 
measuring masses by minimizing the distance to a given isochrone. 

	Note also that I am not using spectral types for the calculation of the IMF. For massive stars, the main advantages of using spectral types to calculate 
IMFs are a better determination of the cluster age, a direct knowledge of the intrinsic color of the star (hence, usually a better extinction measurement),
and an ability to derive additional luminosity/distance information. Those are all possible sources of biases, so spectroscopy indeed represents an 
improvement over a pure photometric analysis. However, as previously stated, for the experiments in this paper I assume that age, extinction, and distance 
are all well constrained by external information, so those advantages of spectroscopy are not relevant for our discussion. The additional advantage that 
spectroscopy has over a simple color-magnitude analysis when blended sources are present is the possibility of identifying them by their composite spectra 
(something that can also be done in some cases with multi-filter photometry). Note, however,
that many spectra of blended sources cannot be distinguished from those of single ones for three possible reasons: [a] need of very high S/N data to detect
the contribution from a dim companion, [b] absence of time-resolved data to detect spectroscopic binaries, and [c] similarity between the two blended 
spectra\footnote{For example, the spectra of the two components of WR 20a \citep{Bonaetal04} are almost indistinguishable and if it were not for the large 
velocity variations induced by their proximity it would be almost impossible to identify the spectrum as a composite.}. In summary, spectroscopy can be and
in most cases is a helpful aid to the elimination of biases in IMF determinations but it also has its costs and limitations in detecting blended sources.
This paper should not be interpreted as neglecting the importance of spectroscopy for the study of the massive-star IMF but rather as a study of what can
be done when spectroscopy is not available.

\subsection{Experiment 1: Real binaries}

	For the first experiment I attempt to reproduce the effect of unresolved multiple systems which, for simplicity, I assume to be always binaries. 
The combination rules applied in this case are:

\begin{enumerate}
  \item Each star is blended with one and only one star.
  \item For a blended system with $m_1\ge m_2$, $m_2$ is randomly selected from a flat distribution in mass between 0.1 \Ms/ (the lower mass limit) 
	and $m_1$. The pairing is done while keeping the real, single-mass stellar IMF as Kroupa (case a) or top-heavy (case b).
\end{enumerate}

\begin{figure}
\centerline{\includegraphics*[width=\linewidth, bb=28 28 566 530]{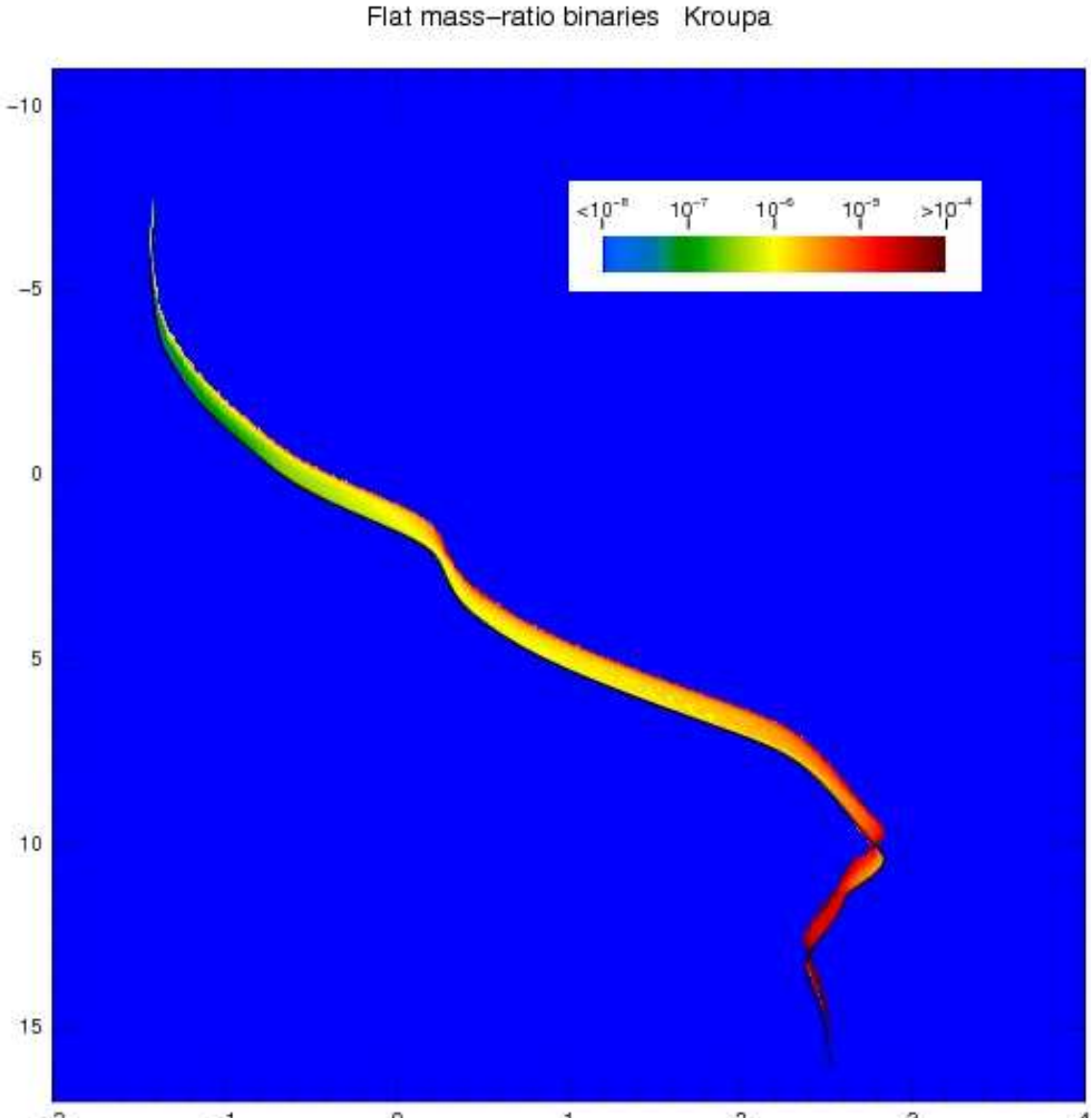}}
\caption{Color-magnitude density function $g_{1{\rm b}}(U-V,V)$ for the Kroupa case of experiment 1 (real binaries) shown as a Hess diagram. The black line shows the 
position of the 1 million year isochrone. The function scaling is logarithmic and the normalization is arbitrary. 
See the electronic version of the journal for a color version of this figure.}
\label{uvv_bin}
\end{figure}	

	This experiment reproduces the physical situation in which all the stars in a cluster are binaries with a flat mass-ratio distribution. As previously 
discussed, this is a rather accurate approximation for systems with at least one massive star but is likely to be more imprecise for lower-mass objects. 
For example, \citet{Reipetal07} observe only an 8.8\% binary fraction among low-mass stars in the Orion Nebula Cluster and \citet{Solletal07} detect a binary
fraction of $\approx$ 40\% for globular clusters with ages less that $10^{10}$ years (of course, the situation in globular clusters is complicated by dynamical
evolution, as the lower fraction detected by the same authors for older clusters show).
However, since our goal is to study the top part of the resulting AMF, the validity of the approximation for low-mass objects should not concern us much. The
other assumption in this experiment is that all massive binaries are unresolved. Under what circumstances is that reasonable? As previously
described, in the solar neighborhood ($\lesssim 2$ kpc) $\sim 1/3$ of the massive systems in clusters/associations are spectroscopic binaries, another 
$\sim 1/3$ are visual/interferometric multiple systems, and the remaining $\sim 1/3$ could be undetected multiples (with periods approximately between those of 
the detected spectroscopic and visual/interferometric systems)\footnote{It should be pointed out that the overlap between spectroscopic and 
visual/interferometric systems is very small, see \citet{Boyaetal07} and \citet{Nortetal07} for examples.}.
This means that even in the solar neighborhood $\sim 2/3$ of the massive binary systems would appear as point sources in a ground-based imaging study with 
standard imaging techniques. Using better spatial resolution (e.g. HST or non-standard techniques such as adaptive optics or lucky imaging) 
buys you a larger fraction of resolved systems in the solar
neighborhood but once you reach several kpc, even that is not enough. For example, ACS/HRC (the imager with the best spatial resolution on HST) can still 
resolve objects with  a magnitude difference $\Delta m \approx 1$ mag and a separation of 50 mas but that is close to the limit of its capabilities 
(see e.g. \citealt{Maizetal08b}). 
At a distance of 7 kpc, 50 mas corresponds to 350 AU, which is the typical value for the separation in the visual O-systems of \citet{Masoetal98}, implying 
that at that distance ACS/HRC would resolve only 20-30\% of the massive binary systems (if their properties are similar to those in the solar neighborhood). The
LMC is hopeless, even with HST: 50 mas corresponds to 2500 AU, so only very-large separation systems are resolved. Therefore, I can conclude that this
experiment represents a reasonable approximation to the conditions in many IMF studies of young clusters and associations beyond the solar neighborhood.

\begin{figure}
\centerline{\includegraphics*[width=.47\linewidth, bb=28 28 566 530]{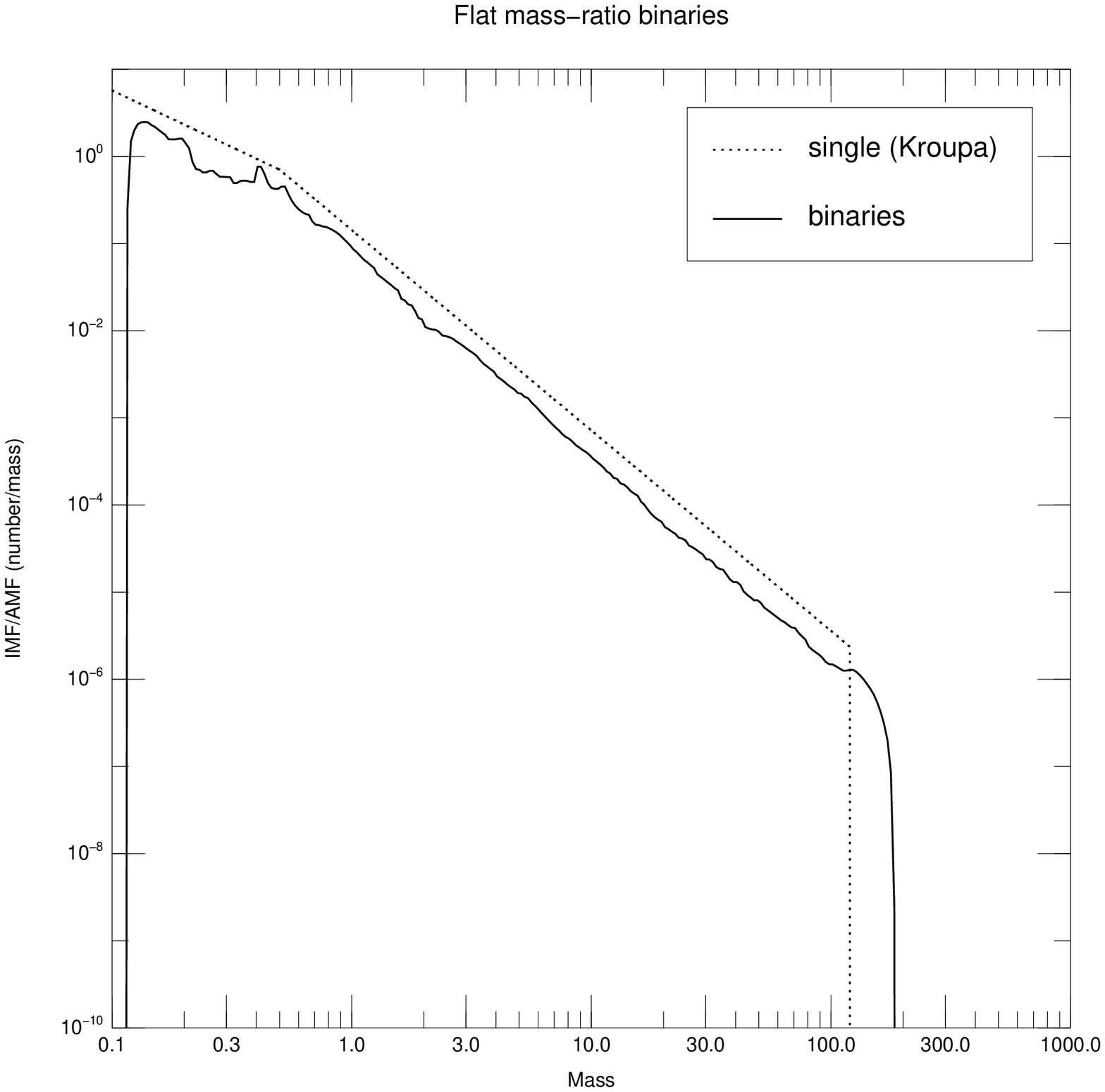}}
\centerline{\includegraphics*[width=.47\linewidth, bb=28 28 566 530]{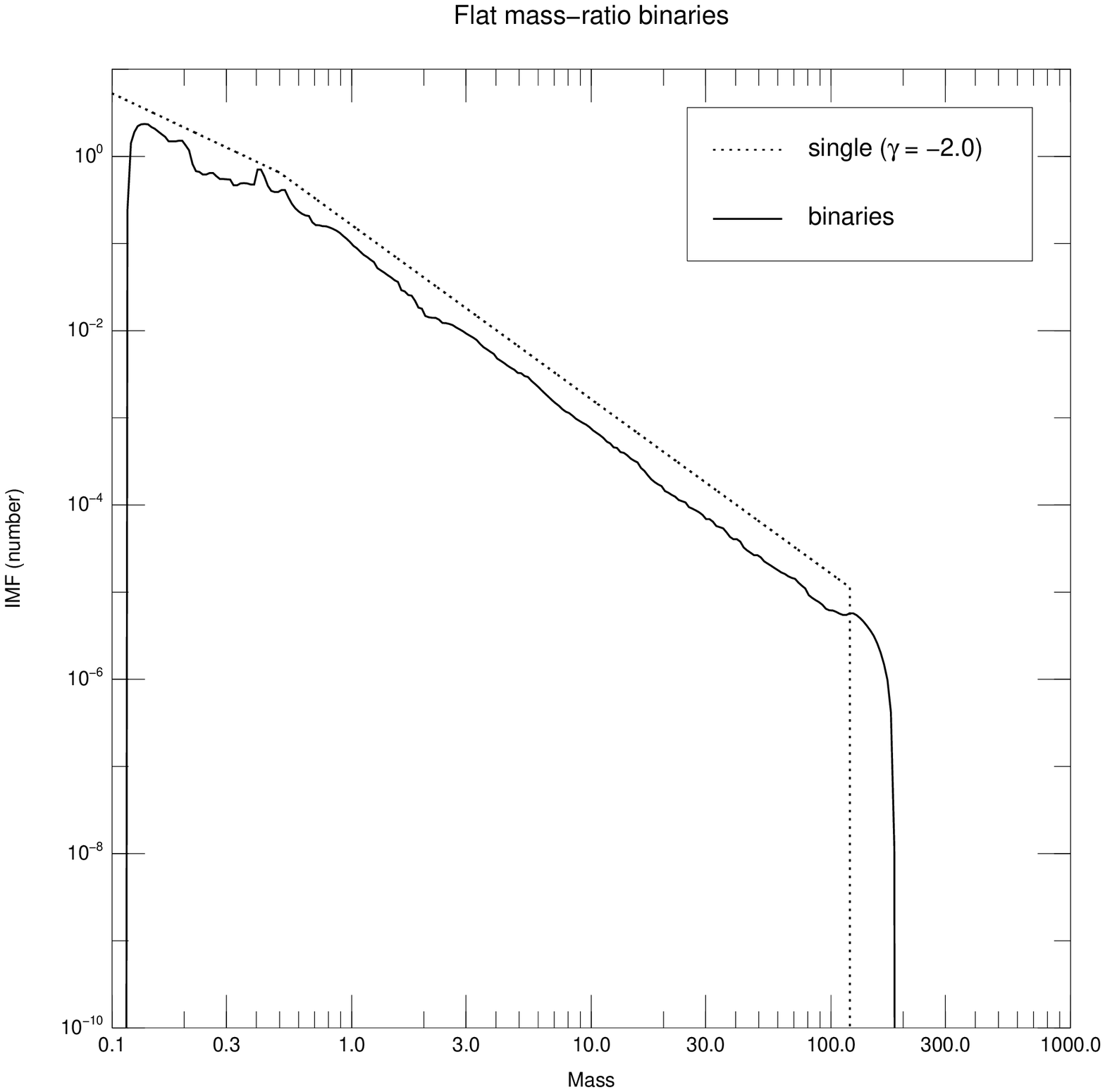}}
\caption{(Real) IMF and AMF for experiment 1 (real binaries). The top panel shows the Kroupa case and the bottom panel the top-heavy case. The mass is
expressed in solar units. The IMF is normalized to 1.0 and the AMF to 0.5.}
\label{imf_bin}
\end{figure}	

	The resulting color-magnitude density function, $g_{1{\rm b}}(U-V,V)$ is shown in Fig.~\ref{uvv_bin} for the Kroupa IMF case (the top-heavy equivalent 
function is not shown but is visually very similar). $g_{1{\rm b}}(U-V,V)$ is zero nearly everywhere in the color-magnitude diagram except in the narrow 
band between the original isochrone and a near-parallel line that is located approximately 0.75 magnitudes above it (the location of the equal-mass binaries). The band 
between the two lines is filled with the non-equal-mass binaries. At the low-mass extreme the band becomes narrower and detached from the isochrone because the 
addition of any companion changes the magnitude (and possibly the color) considerably. That effect does not take place for higher masses because the addition 
of a 0.1 \Ms/ star is undetectable in color-magnitude with the scale of Fig.~\ref{uvv_bin}. At the high-mass end an obvious extension beyond $m_{\rm max}$
is observed. The extension is nearly vertical because massive stars of this age have near-degenerate $U-V$ colors. The apparent upper mass limit, $m_{{\rm max},1{\rm b}}$ is
well defined and has a value of 182 \Ms/.

\begin{figure}
\centerline{\includegraphics*[width=\linewidth, bb=28 28 566 535]{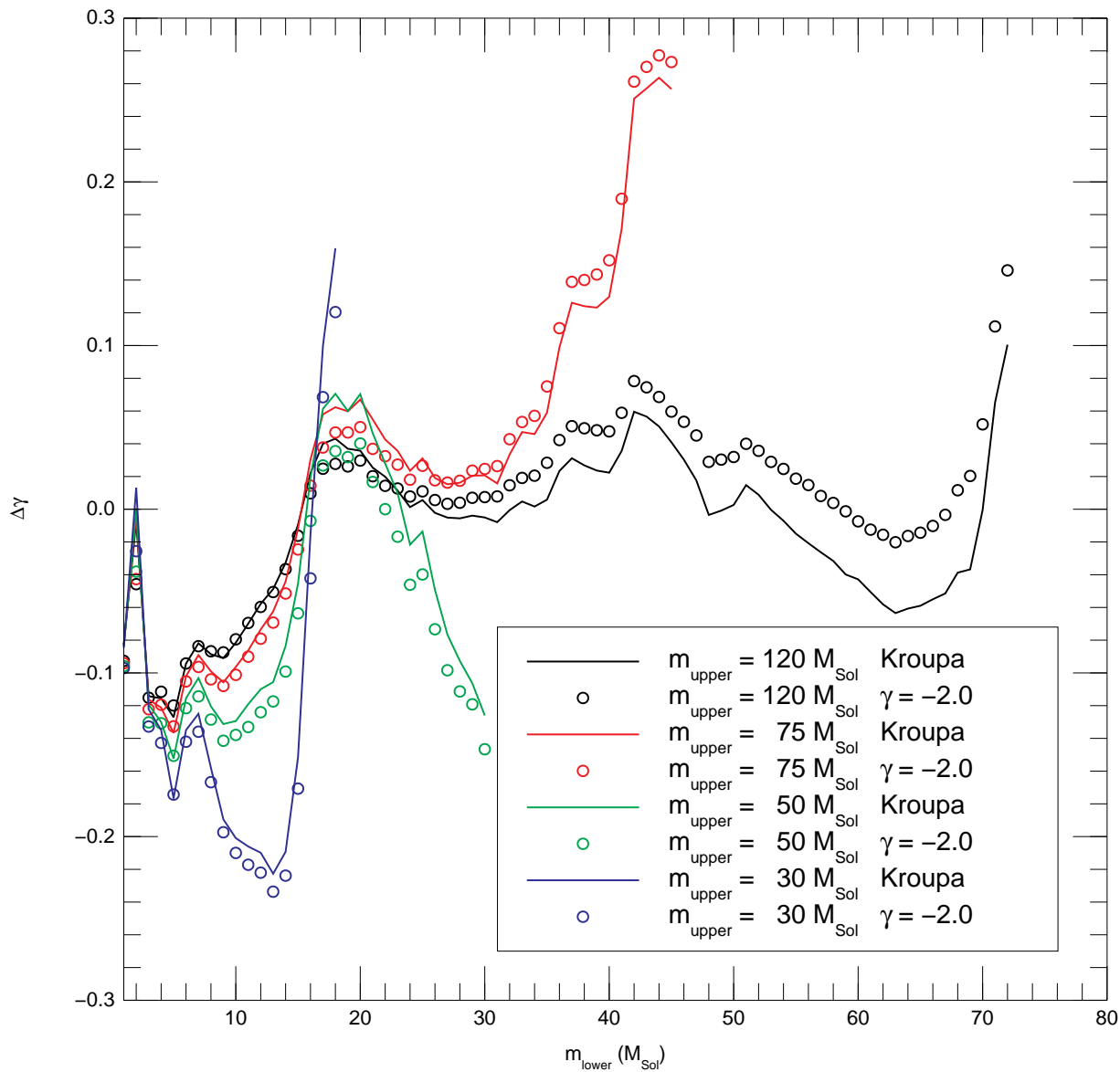}}
\caption{Slope change introduced by flat mass-ratio binaries (first experiment) as a function of the lower limit for the mass range ($m_{\rm lower}$) for the two 
IMF cases and for different values of the upper limit for the mass range ($m_{\rm upper}$). In each case the values between 1 \Ms/ and 0.60$m_{\rm upper}$ are plotted. 
See the electronic version of the journal for a color version of this figure.}
\label{deltaslope_bin}
\end{figure}	

\begin{deluxetable}{ccc}
\tablecaption{Apparent ratios between the number of objects above 120 \Ms/ and the number in different mass ranges for the two cases of the first experiment.}
\tablewidth{0pt}
\tablehead{Mass range & Kroupa   & Top-heavy \\
           \Ms/       & real IMF & real IMF   }
\startdata
       All &  $ 7.873\cdot 10^{-5}$ &  $ 3.639\cdot 10^{-4}$ \\
 $  1-120$ &  $ 6.661\cdot 10^{-4}$ &  $ 2.246\cdot 10^{-3}$ \\
 $  3-120$ &  $ 2.835\cdot 10^{-3}$ &  $ 7.142\cdot 10^{-3}$ \\
 $  5-120$ &  $ 5.941\cdot 10^{-3}$ &  $ 1.294\cdot 10^{-2}$ \\
 $  8-120$ &  $ 1.164\cdot 10^{-2}$ &  $ 2.225\cdot 10^{-2}$ \\
 $ 15-120$ &  $ 2.921\cdot 10^{-2}$ &  $ 4.726\cdot 10^{-2}$ \\
 $ 25-120$ &  $ 6.206\cdot 10^{-2}$ &  $ 8.895\cdot 10^{-2}$ \\
 $ 40-120$ &  $ 1.284\cdot 10^{-1}$ &  $ 1.662\cdot 10^{-1}$ \\
 $ 60-120$ &  $ 2.656\cdot 10^{-1}$ &  $ 3.163\cdot 10^{-1}$ \\
\enddata
\label{ratios_bin}
\end{deluxetable}

	The calculated AMFs for the Kroupa and top-heavy cases are shown in Fig.~\ref{imf_bin} along with the original IMFs. The behavior in both cases is
similar. For low masses, the apparent lower limit is slightly shifted towards higher values (corresponding to a 0.1~\Ms/~+~0.1~\Ms/ binary, since 0.1~\Ms/
is the lower mass limit of our experiments)
and several peaks in the AMF can be seen below 0.7 \Ms/ due to the fine structure of $m(V)$\footnote{The conversion from a magnitude or luminosity function
to a IMF involves $dm/dV$, whose numerical behavior may be odd when using coarsely-gridded tabular information to derive it. The observed structures may or may
not be real but since our concern is with the high-mass end of the AMF, we will not discuss them any further.}. For intermediate and high masses, the AMF runs 
nearly parallel to the IMF at a value roughly 1/2 of the IMF (the AMF includes one half of the systems of the IMF). In order to quantify the change in slope 
between the IMF and the binary AMF, $\Delta\gamma$, I have fitted a power law to a number of mass ranges using the technique in Paper I.
Results are plotted in Fig.~\ref{deltaslope_bin}. For $m_{\rm lower}$ between 1 and 16 \Ms/ and $m_{\rm upper} \ne$ 120 \Ms/ 
there is a steepening of the MF with $\Delta\gamma$ between $-0.2$ and $0.0$ while for larger masses in most cases there a slight flattening with 
$\Delta\gamma$ between $0.0$ and $0.2$. $\Delta\gamma$ depends only weakly on $m_{\rm upper}$ and the IMF slope.
Note that some of the small-scale structure observed in Fig.~\ref{deltaslope_bin} at the 0.02 level may be of 
numerical origin. I conclude that for most cases {\it the existence of unresolved binaries has only a small effect on the massive-star IMF slope}. 

	The largest difference between the IMF and the AMF happens beyond $m_{\rm max}$, where a tail is formed in the AMF: these are the apparently
ultramassive stars (AUMSs). In Table~\ref{ratios_bin} I quantify the effect by giving the apparent number ratios between the
objects above $m_{\rm max}$ and the objects in different ranges below it. There is a strong dependence on the IMF slope,
since a top-heavy IMF allows a significantly larger number of objects to become AUMSs.

	Figure~\ref{uvv_bin} reveals another interesting issue which is relevant to IMF studies of Galactic clusters, where distance is one of the unknowns that
has to be solved for. O stars are notoriously difficult to use as yardsticks because their near-degenerate optical colors place them along near-vertical lines in a 
color-magnitude diagram (see e.g. Fig.~\ref{uvv_bin} but see also \citealt{MaizSota08} for recent progress on improving our knowledge of O-stars colors). For that
reason, main-sequence B stars (which have intrinsic $U-V$ in the approximate range from $-1.3$ to $-0.1$) are more commonly used to derive distances with main-sequence 
fitting techniques. However, if the assumption which underlies our experiment is relevant (that most O and B stars are unresolved binaries with flat mass-ratio 
distributions), then it is easy to see that a direct application of main-sequence fitting to a $U-V$ vs. $V$ diagram would lead to an overestimation of the distance, 
because for a fixed value of $U-V$ there should be more stars close to the equal-mass binary sequence than to the main sequence itself. Note that this effect should 
be significantly smaller for older clusters because the binary fraction is lower there, as it is indeed observed for globular clusters (see e.g. \citealt{Solletal07}). 
Spectroscopy can partially resolve the problem by detecting blended spectra or radial velocity variations but from Fig.~\ref{uvv_bin} we can see that it should be
also possible to correct for the effect by considering that there is an intrinsic spread in the observed color-magnitude diagram due to binarity (besides the 
ones due to differential extinction and foreground/background contamination).

\subsection{Experiment 2: Chance superpositions of single stars}

	In the second experiment I model the effect of chance superpositions between single (not binary) stars within the observed clusters. Such a simulation
introduces additional complications that are detector- and observation-dependent because now we have to deal with a continuous spatial distribution of stars along a 
2-D detector. In other words, the observed luminosity function depends not only on the intrinsic properties of the cluster (e.g. its internal structure) but also on the
the relative pixel and point-spread function (PSF) sizes, Poisson statistics, and the ability of the detector and the finding algorithm to differentiate between single- 
and multiple- point sources of diverse $\Delta m$. Given those considerations, the goal in this case is to obtain a toy model that allows us to 
approximately quantify the effect for a given region of a cluster in order to test the magnitude of the effect.

	The combination rules in this case are:

\begin{enumerate}
  \item The light from a given star falls in a single pixel (the {\it fat pixel approximation}).
  \item The initial cluster is built by generating a fine grid in $U-V$ and $V$ and assigning to each point a value of $g_{1{\rm s}}(U-V,V)$ integrated over the size 
	of the bin. Since $g_{1{\rm s}}(U-V,V)$ is zero outside the 1-million year isochrone, most points in the grid have no stars. The initial cluster is then
	defined by the non-zero values of $g_{1{\rm s}}(U-V,V)$. As with the previous experiment, we consider the two cases of a Kroupa and a top-heavy IMF.
  \item $g_{2{\rm s}}(U-V,V)$ is a color-magnitude function that describes a cluster with two single stars in each pixel which are randomly paired. I generate it by
	convolving $g_{1{\rm s}}(U-V,V)$ with itself and assigning the result of each pair to the closest value in the ($U-V,V$) grid\footnote{As previously noted, 
	special care was taken to avoid roundoff errors.}. Note that $g_{2{\rm s}}(U-V,V)$ does not take into consideration Poisson statistics for the stellar 
	distribution among pixels but instead considers that each pixel contains exactly two stars.
  \item $g_{4{\rm s}}(U-V,V)$ is generated by convolving $g_{2{\rm s}}(U-V,V)$ with itself, and in general, $g_{2^ns}(U-V,V)$ is generated by convolving 
        $g_{2^{n-1}{\rm s}}(U-V,V)$ with itself. In this paper I generate the color-magnitude functions up to $n=11$ i.e. $g_{2048{\rm s}}(U-V,V)$.
\end{enumerate}

	Those rules should be able to simulate the general behavior of the observed color-magnitude diagram as the intrinsic stellar density or the cluster distance
increases. However, there are three caveats to its direct application. The first one is that a comparison with a real case should always consider the possible
additional effect of Poisson fluctuations, especially when the number of superimposed stars $N_{\rm sup}=2^n$ is small. 
One way to do this if the expected average number of stars per pixel, $N_{\rm px}$,
falls between $2^{n-1}$ and $2^n$ would be to consider both of those cases as possible. The second caveat is that the fat pixel approximation is not valid for a
well-sampled detector (PSF size several times larger than the pixel size), so an effective pixel size has to be defined by calculating the area of influence of a star
i.e. the region where the presence of a star does not allow a second star to be detected but instead the fluxes from both are merged. Such an effective pixel size is
in reality $\Delta m$-dependent so an average effective pixel size should be calculated. Also, the precise flux derived depends on the photometric technique (PSF fitting
or aperture photometry) and the algorithm details. Once again, I stress that the purpose of this experiment is not to find the
exact correction required to account for chance superpositions but rather to develop a toy model that would allow the observer to roughly quantify the effect. The third
caveat is that all stars are assumed to be single. This last caveat will be addressed by the third experiment.

\begin{figure}
\centerline{\includegraphics*[width=\linewidth, bb=130 505 475 720]{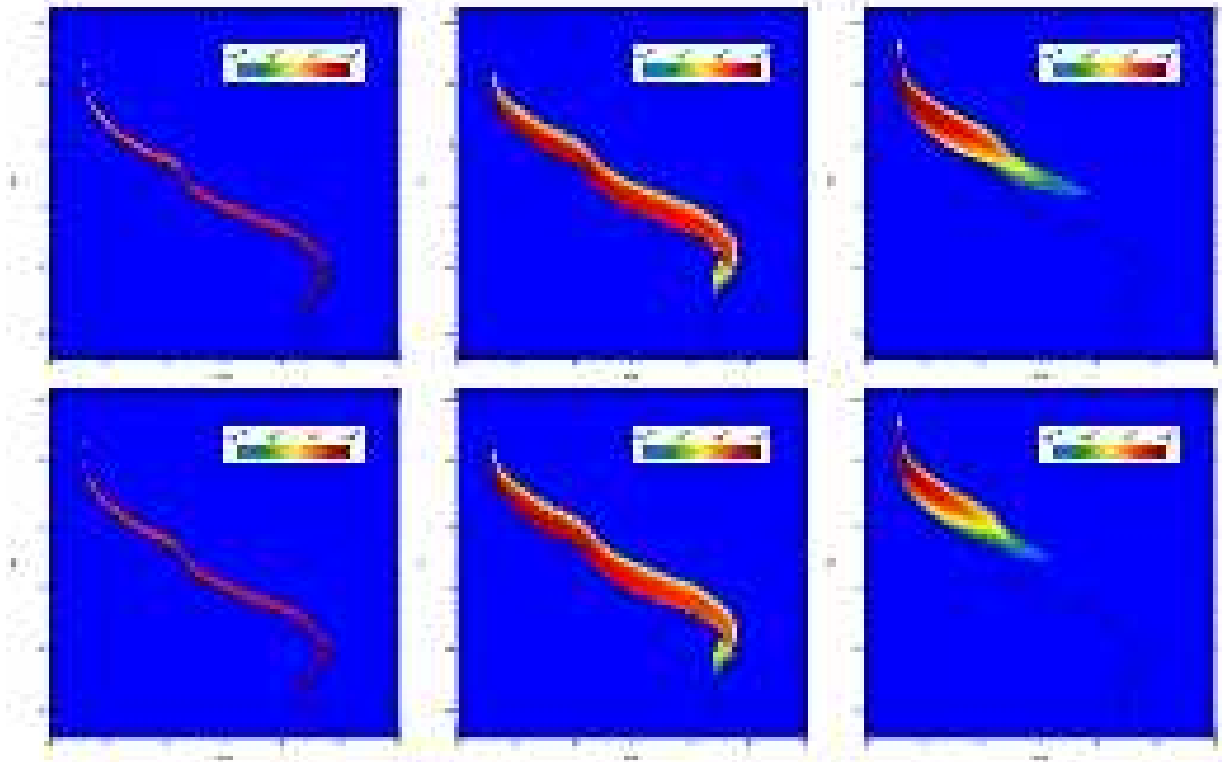}}
\caption{Color-magnitude density functions $g_{2{\rm s}}(U-V,V)$ (left), $g_{16{\rm s}}(U-V,V)$ (center), and $g_{128{\rm s}}(U-V,V)$ (right) for the Kroupa (top row)
and top-heavy (bottom row) cases of experiment 2 (chance superpositions of single stars) shown as Hess diagrams. The thick black line shows the 
position of the 1 million year isochrone. The function scaling is logarithmic and the normalization is arbitrary.
See the author's web site {\tt http://www.iaa.es/$_{\mbox{\~{}}}$jmaiz} for animated gifs showing all the color-magnitude functions $g_{2^n{\rm s}}(U-V,V)$ from
$n=0$ to $n=11$.
See the electronic version of the journal for a color version of this figure.}
\label{uvv_chsup1}
\end{figure}	

\begin{figure}
\centerline{\includegraphics*[angle=90, width=\linewidth]{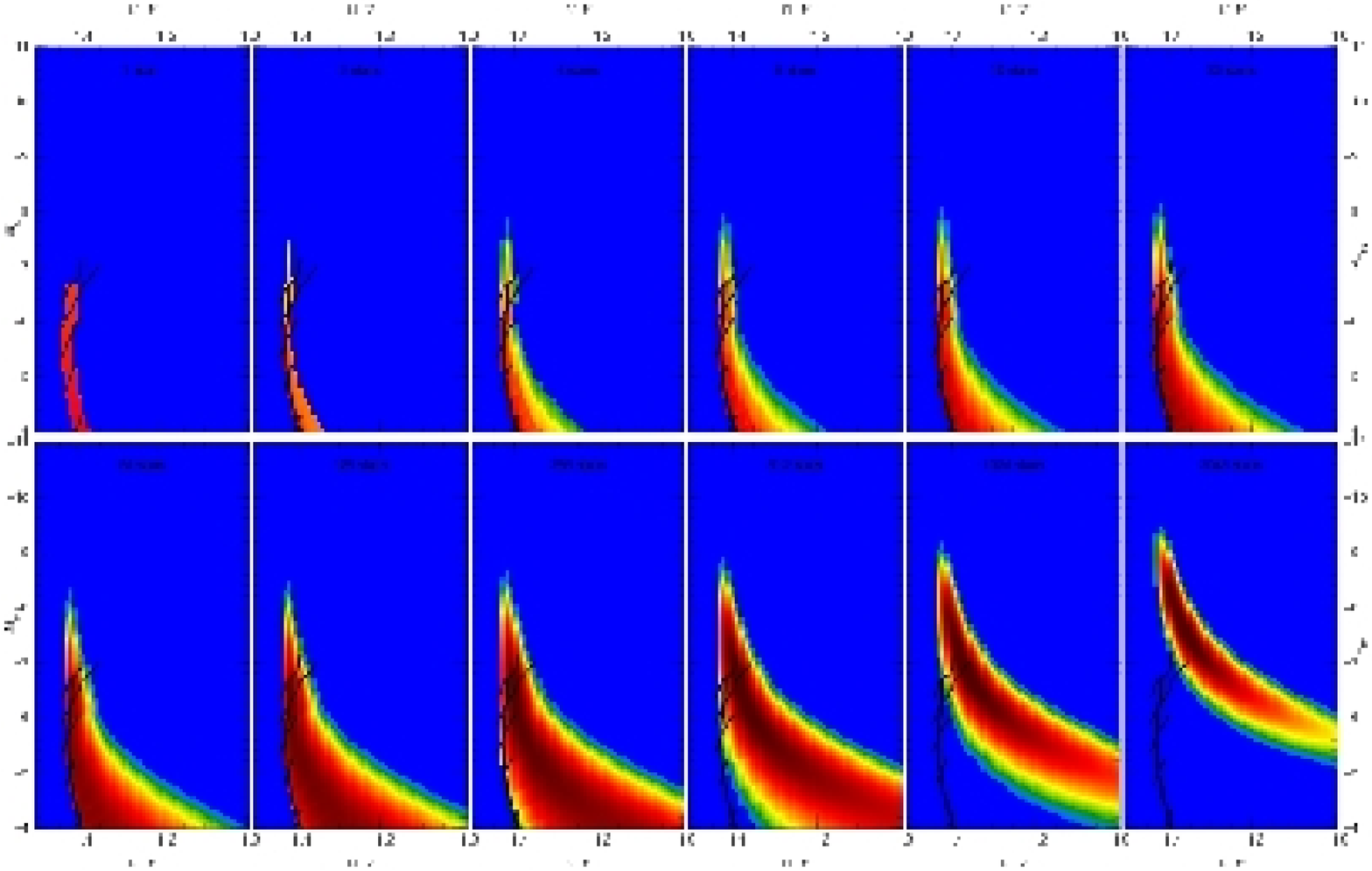}}
\caption{Top left corner of the color-magnitude density functions $g_{2^n{\rm s}}(U-V,V)$ with $n=0,11$ for the Kroupa case of experiment 2 (chance superpositions
of single stars) shown as Hess diagrams. The $n=0$ case has been smoothed in order to allow it to show some extent perpendicular to the 1 million year isochrone.
The thick black line shows the position of the 1 million year isochrone and the thin lines the evolutionary tracks between 0 and 2 million years for the initial masses
of 25, 40, 60, 85, and 120 \Ms/. The function scaling is logarithmic and the normalization is arbitrary.
See the electronic version of the journal for a color version of this figure.}
\label{uvv_chsup2}
\end{figure}	

\begin{figure}
\centerline{\includegraphics*[angle=90, width=\linewidth]{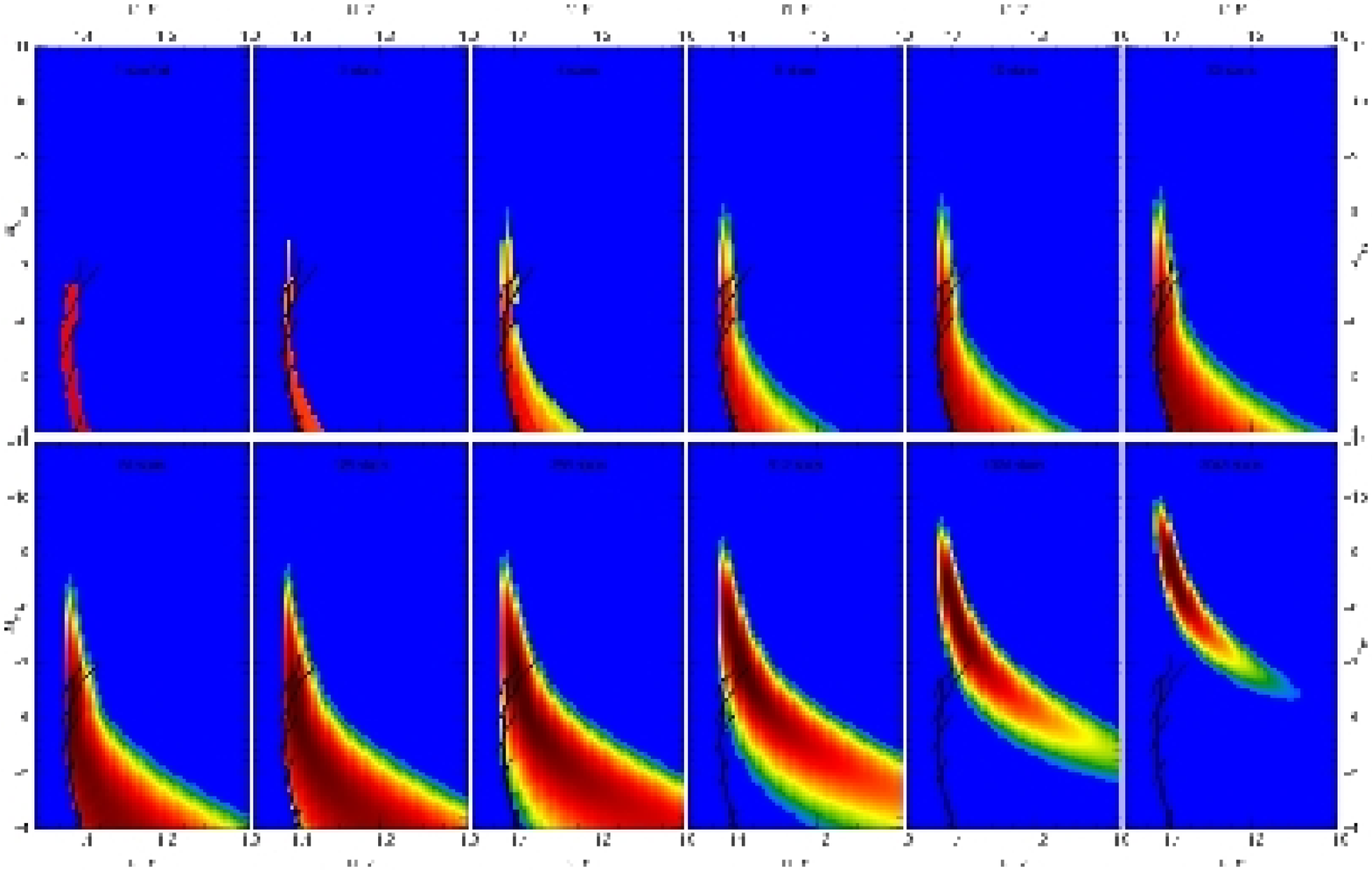}}
\caption{Top left corner of the color-magnitude density functions $g_{2^n{\rm s}}(U-V,V)$ with $n=0,11$ for the top-heavy case of experiment 2 (chance superpositions
of single stars) shown as Hess diagrams. The $n=0$ case has been smoothed in order to allow it to show some extent perpendicular to the 1 million year isochrone.
The thick black line shows the position of the 1 million year isochrone and the thin lines the evolutionary tracks between 0 and 2 million years for the initial masses
of 25, 40, 60, 85, and 120 \Ms/. The function scaling is logarithmic and the normalization is arbitrary.
See the electronic version of the journal for a color version of this figure.}
\label{uvv_chsup3}
\end{figure}	

	Six of the resulting full color-magnitude density functions, $g_{2^n{\rm s}}(U-V,V)$, are shown in Fig.~\ref{uvv_chsup1} for $n=1,4,7$ and the
two cases (Kroupa and top heavy). The top left corner (where massive stars are located) of all the functions is shown in Figs.~\ref{uvv_chsup2} (Kroupa) and
\ref{uvv_chsup3} (top heavy). As was the case in experiment 1, $g_{2^n{\rm s}}(U-V,V)$ is zero nearly everywhere in the color-magnitude diagram except in a narrow 
band. The band widens up until $n=8$ and then narrows down from that point onwards. 
At the same time, the lower end of the band starts climbing up the vertical coordinate starting at $n=1$.
The top end starts climbing up at a lower pace but then picks up speed around $n=8$. The progression of both ends is faster for the top-heavy case than for the
Kroupa case. For the final $n=11$ value the resulting Hess-diagram has been compressed to a well-defined narrow peak in the $U-V$ vs. $V$ color diagram: the
distribution is completely contained in the last plot of the top-heavy case of Fig.~\ref{uvv_chsup3} and almost so for the Kroupa case of 
Fig.~\ref{uvv_chsup2}.

	Figures~\ref{uvv_chsup1},~\ref{uvv_chsup2},~and~\ref{uvv_chsup3} indicate that, as crowding becomes more of an issue, we progressively lose the ability to 
detect low-mass stars. Once a certain threshold has been attained, high-mass stars also become significantly affected by crowding. However, one should be careful in
overinterpreting those Hess diagrams as color-magnitude probability distributions for the observed population of points source in a a given cluster because of
the fat pixel assumption. The lower end of the observed band represents in reality those pixels which detection algorithms would simply classify as background.
Therefore, a more realistic interpretation of the Hess diagrams would be that the observed point source population measured by assuming a zero background (or
calculating it outside the cluster boundaries) is extracted from the top part of the 2-D function given by $g_{2^n{\rm s}}(U-V,V)$ and that the lower part
represents the unobserved part of the population. The limit between the two regions is defined by the characteristics of the detector and the observations. 

	An interesting alternative interpretation of Figs.~\ref{uvv_chsup1},~\ref{uvv_chsup2},~and~\ref{uvv_chsup3} is that they represent the probability
distribution for a cluster with $2^n$ stars, with the extension in $U-V$ and $V$ indicating the result of stochastic effects in drawing the given number of stars from
an IMF. I will not explore this interpretation any further in this article; the reader is referred to \citet{CervLuri06} for more details.

	It is also useful to compare the results from $g_{1{\rm b}}$ (first experiment) and $g_{2{\rm s}}$ (second experiments)\footnote{Note that the function scaling
in Figs.~\ref{uvv_bin}~and~\ref{uvv_chsup1} is different.}. Both functions represent the combination of two stars but with an important difference: in the first case 
the pairing assumes a flat mass-ratio distribution while in the second case the pairing is random. The different combination rules manifest themselves in the slope that
is obtained when plotting $g_{1{\rm b}}$ or $g_{2{\rm s}}$ as a function of $V$ for a fixed $U-V$. For most values of $U-V$ in the first case the function increases
towards lower values of $V$ (higher luminosities) while in the second one the behavior is the opposite. The reason is that for random pairing, once an $m_1$ is 
drawn in the middle or upper main sequence, chances are that $m_2$ will be significantly lower. On the other hand, for a flat mass-ratio distribution one gets a larger 
contribution from stars with $m_2$ only slightly lower than $m_1$, thus tilting the balance at a fixed $U-V$ towards systems with near-equal mass ratios.

\begin{figure}
\centerline{\includegraphics*[width=0.61\linewidth, bb=155 175 450 720]{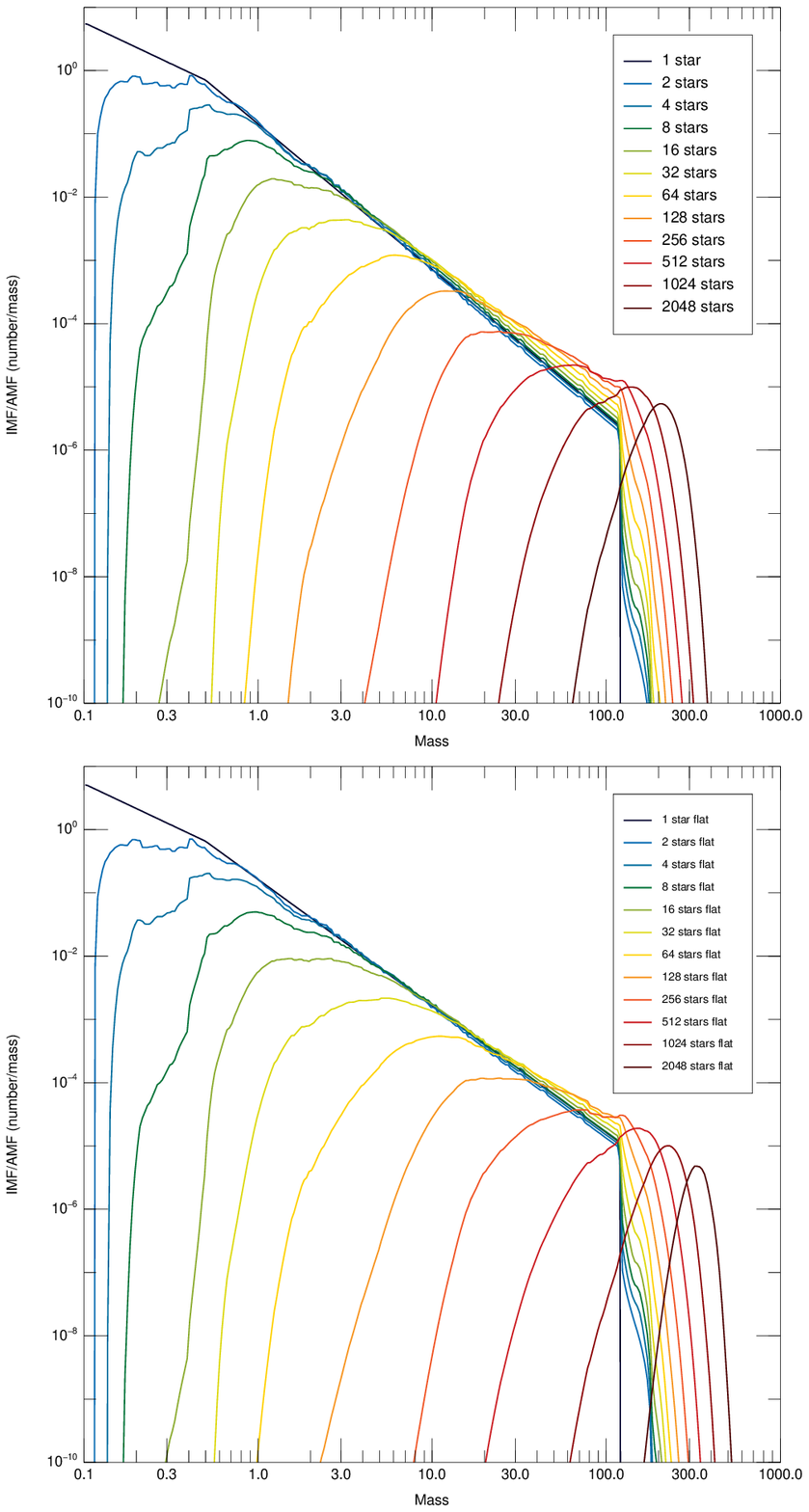}}
\caption{IMF and AMFs for experiment 2 (chance superpositions of single stars). The top panel shows the Kroupa case and the bottom panel the top-heavy case. The mass is
expressed in solar units. The IMF is normalized to 1.0 and the AMFs to $2^{-n}$.}
\label{imf_chsup}
\end{figure}	

\begin{deluxetable}{ccccc}
\tablecaption{Apparent ratios between the number of objects above 120 \Ms/ and the total number of objects for the second (single stars) and third (binaries) experiments.}
\tablewidth{0pt}
\tablehead{\# of   & Kroupa  & Top-heavy & Kroupa   & Top-heavy \\
           objects & singles & singles   & binaries & binaries  }
\startdata
   1 &  $ 0.000\cdot 10^{+0}$ &  $ 0.000\cdot 10^{+0}$ &  $ 7.873\cdot 10^{-5}$ &  $ 3.639\cdot 10^{-4}$ \\
   2 &  $ 2.080\cdot 10^{-7}$ &  $ 3.331\cdot 10^{-6}$ &  $ 1.913\cdot 10^{-4}$ &  $ 8.708\cdot 10^{-4}$ \\
   4 &  $ 1.027\cdot 10^{-6}$ &  $ 1.582\cdot 10^{-5}$ &  $ 4.458\cdot 10^{-4}$ &  $ 1.988\cdot 10^{-3}$ \\
   8 &  $ 4.829\cdot 10^{-6}$ &  $ 7.101\cdot 10^{-5}$ &  $ 1.033\cdot 10^{-3}$ &  $ 4.552\cdot 10^{-3}$ \\
  16 &  $ 2.485\cdot 10^{-5}$ &  $ 3.552\cdot 10^{-4}$ &  $ 2.469\cdot 10^{-3}$ &  $ 1.082\cdot 10^{-2}$ \\
  32 &  $ 1.422\cdot 10^{-4}$ &  $ 1.977\cdot 10^{-3}$ &  $ 6.209\cdot 10^{-3}$ &  $ 2.703\cdot 10^{-2}$ \\
  64 &  $ 9.067\cdot 10^{-4}$ &  $ 1.155\cdot 10^{-2}$ &  $ 1.646\cdot 10^{-2}$ &  $ 7.090\cdot 10^{-2}$ \\
 128 &  $ 6.040\cdot 10^{-3}$ &  $ 6.253\cdot 10^{-2}$ &  $ 4.626\cdot 10^{-2}$ &  $ 1.948\cdot 10^{-1}$ \\
 256 &  $ 3.697\cdot 10^{-2}$ &  $ 2.790\cdot 10^{-1}$ &  $ 1.385\cdot 10^{-1}$ &  $ 5.045\cdot 10^{-1}$ \\
 512 &  $ 1.907\cdot 10^{-1}$ &  $ 7.765\cdot 10^{-1}$ &  $ 4.079\cdot 10^{-1}$ &  $ 9.118\cdot 10^{-1}$ \\
1024 &  $ 6.667\cdot 10^{-1}$ &  $ 9.979\cdot 10^{-1}$ &  $ 8.684\cdot 10^{-1}$ &  $ 9.998\cdot 10^{-1}$ \\
2048 &  $ 9.944\cdot 10^{-1}$ &  $ 1.000\cdot 10^{+0}$ &  $ 9.997\cdot 10^{-1}$ &  $ 1.000\cdot 10^{+0}$ \\
\enddata
\label{ratios_chsup}
\end{deluxetable}

\begin{figure}
\centerline{\includegraphics*[width=\linewidth, bb=28 28 566 535]{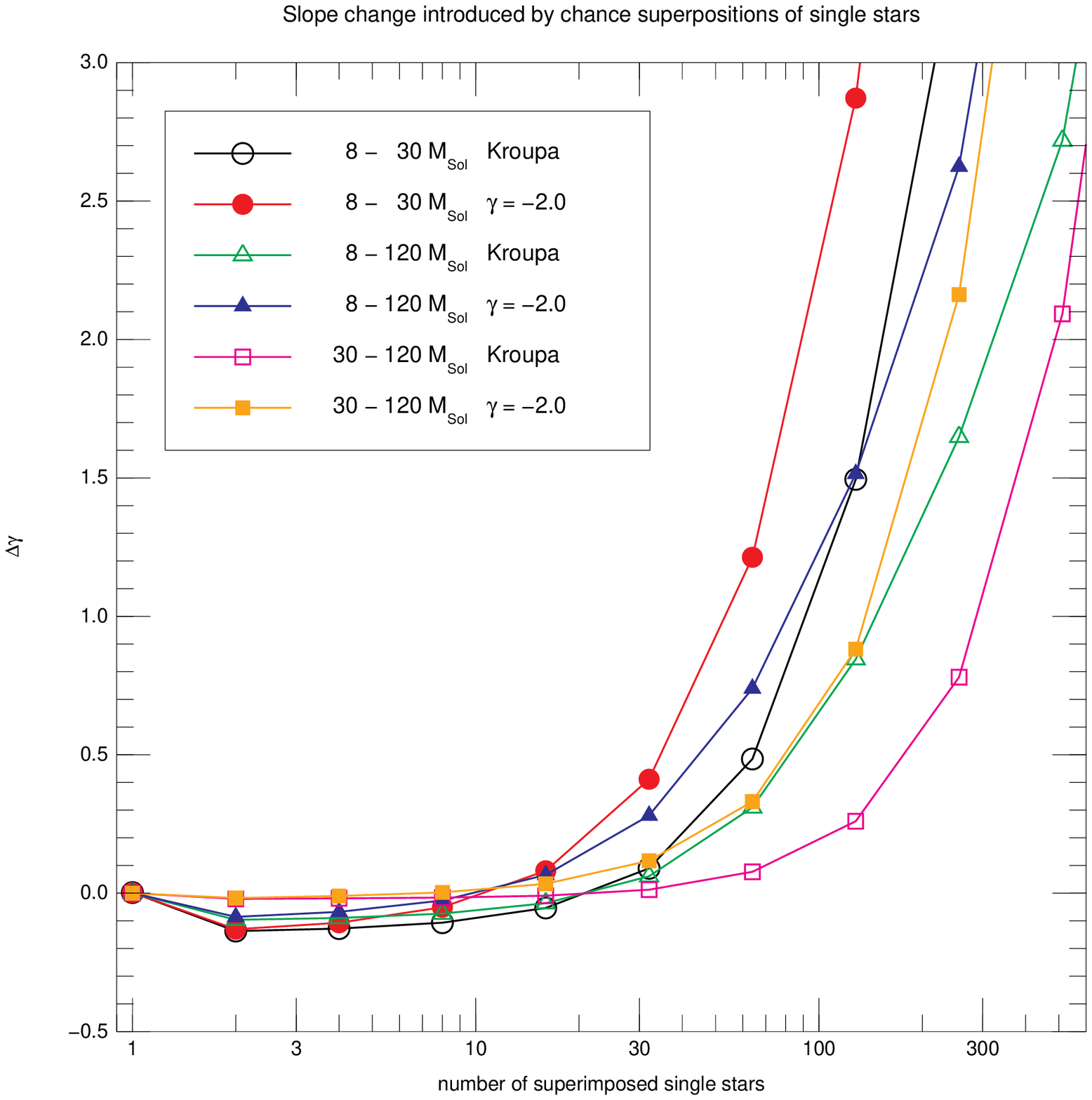}}
\caption{Slope change introduced by chance superpositions of single stars (second experiment) as a function of the number of superimposed single stars for the two 
IMF cases and for three different mass ranges.
See the electronic version of the journal for a color version of this figure.}
\label{deltaslope_chsup}
\end{figure}	

	The calculated AMFs for the Kroupa and top-heavy cases are shown in Fig.~\ref{imf_chsup} along with the original IMFs. The overall behavior in both
cases again being similar: as the number of stars increases, the left extreme of the AMF moves towards the right in approximately equal amounts in $\log m$ for each 
step in $n$. The right extreme has a different behavior: the population of AUMSs initially appears only as a weak tail and it is only when
$n=7-8$ that a significant number is built up. From then on, the right extreme of the AMF picks up speed and starts moving towards the right, though at a lower pace
than the left extreme. As a consequence, the distribution narrows in $\log m$. The narrowing process also affects the overall shape of the distribution, which for
$n=11$ is already quite similar to a log-normal (especially for the top-heavy case).

	Table~\ref{ratios_chsup} lists the fraction of objects with apparent masses above $m_{\rm max}$ = 120 \Ms/ 
for both the Kroupa (second column) and top-heavy cases (third column) as a 
function of $N_{\rm sup}$. The fraction is highly dependent on the IMF slope, with the top-heavy case showing values an order of magnitude above the Kroupa case until the point
where the value becomes close to unity for the top-heavy case ($n\approx 9$). Such a dependence is not an unexpected behavior, because even a small change in $\gamma$ is
sufficient to produce a large variation in the ratio of stars with masses around 100 \Ms/ to low-mass stars in the real IMF (compare e.g. the two panels in Fig.~\ref{imf_bin}).
As also expected, flat-mass ratio binaries are significantly more efficient than chance superpositions in creating a population of AUMSs: $\approx 32$ 
superimposed stars are required to create the same number of those stars as those created by a single binary for the Kroupa case (the number is slightly lower for the 
top-heavy case).

	Figure~\ref{deltaslope_chsup} plots $\Delta\gamma$ the change in MF slope as a function of $N_{\rm sup}$ for three mass ranges. Three regimes can be distinguished. For 
low $N_{\rm sup}$, the effect in $\Delta\gamma$ is small and in most cases implies a slight steepening of order 0.1. For intermediate values, $\Delta\gamma$ starts to increase
(the mass function becomes flatter) but the effect remains small and should be correctable. Finally, for large $N_{\rm sup}$, $\Delta\gamma$ becomes so large that no realistic
correction is possible. 

\subsection{Experiment 3: Chance superpositions of binaries}

	The third experiment is a combination of the previous two. I start with the color-magnitude generated in the first experiment for flat mass-ratio binaries,
$g_{1{\rm b}}(U-V,V)$, and I convolve it with itself to derive $g_{2{\rm b}}(U-V,V)$. $g_{4{\rm b}}(U-V,V)$ is then generated by convolving $g_{2{\rm s}}(U-V,V)$ with itself
and the process is repeated until $n=11$ i.e. $g_{2048{\rm b}}(U-V,V)$. This experiment should provide the most realistic representation of the observed color-magnitude density
function and AMF of a distant cluster, always keeping in mind the first two caveats previously discussed.

\begin{figure}
\centerline{\includegraphics*[width=\linewidth, bb=130 505 475 720]{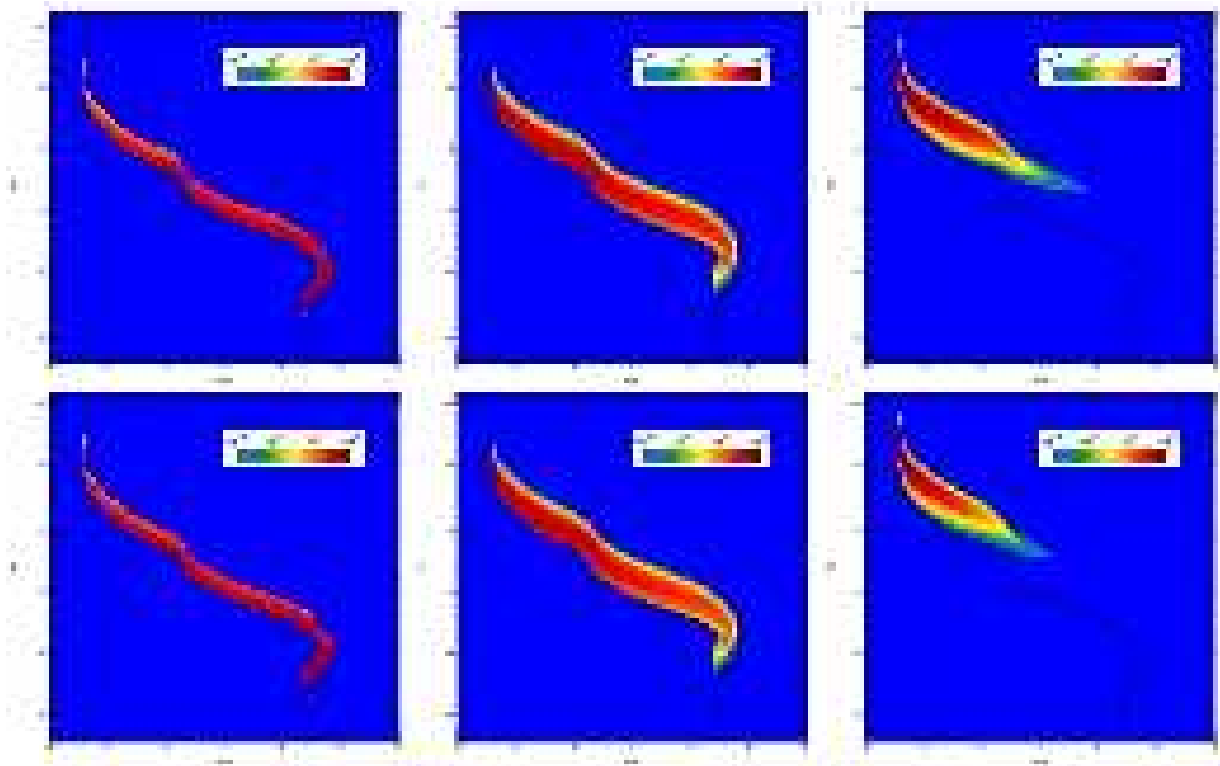}}
\caption{Color-magnitude density functions $g_{2{\rm b}}(U-V,V)$ (left), $g_{16{\rm b}}(U-V,V)$ (center), and $g_{128{\rm b}}(U-V,V)$ (right) for the Kroupa (top row)
and top-heavy (bottom row) cases of experiment 3 (chance superpositions of binaries) shown as Hess diagrams. The thick black line shows the 
position of the 1 million year isochrone. The function scaling is logarithmic and the normalization is arbitrary.
See the author's web site {\tt http://www.iaa.es/$_{\mbox{\~{}}}$jmaiz} for animated gifs showing all the color-magnitude functions $g_{2^n{\rm b}}(U-V,V)$ from
$n=0$ to $n=11$.
See the electronic version of the journal for a color version of this figure.}
\label{uvv_chsupbin1}
\end{figure}	

\begin{figure}
\centerline{\includegraphics*[angle=90, width=\linewidth]{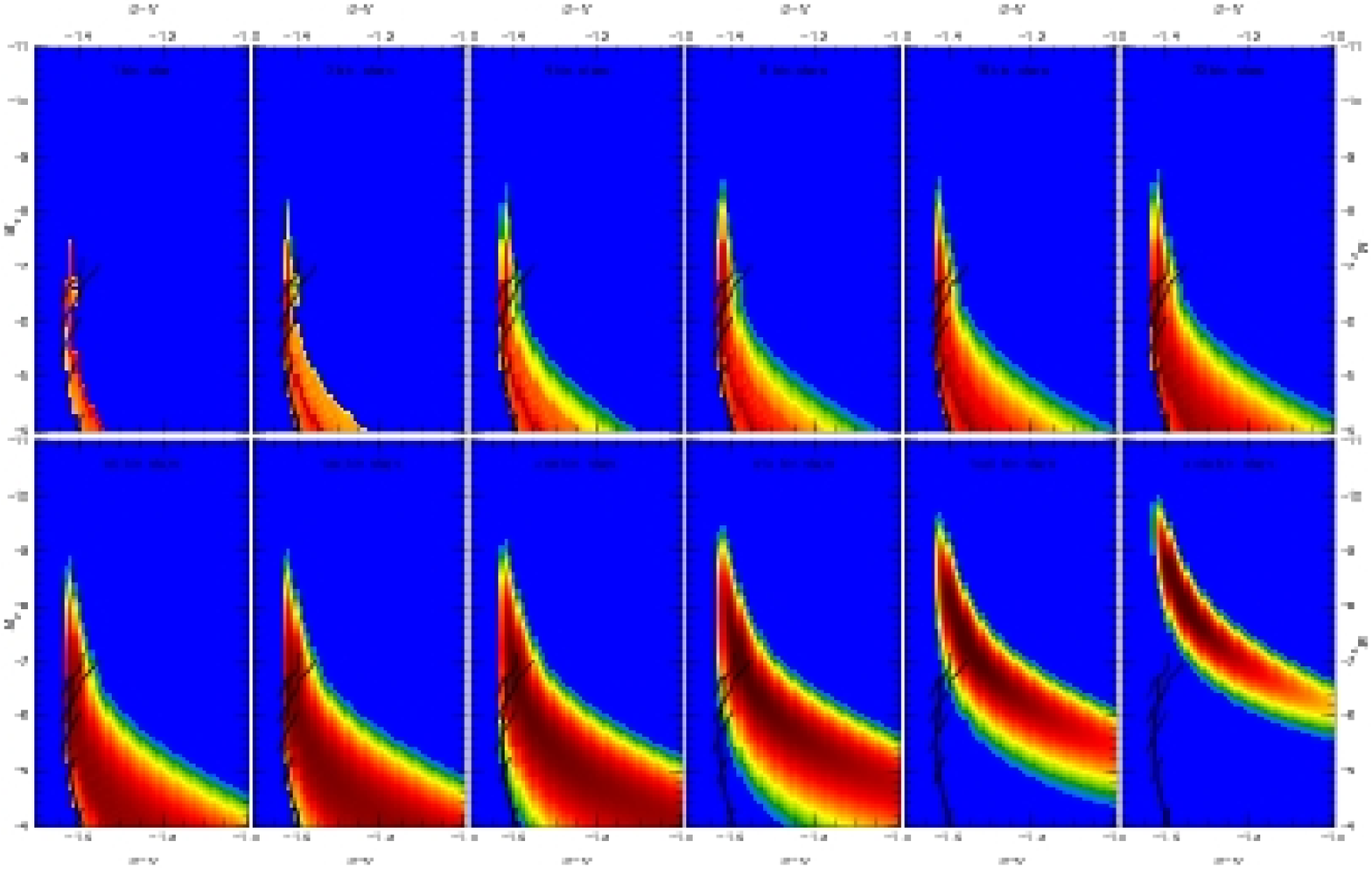}}
\caption{Top left corner of the color-magnitude density functions $g_{2^n{\rm b}}(U-V,V)$ with $n=0,11$ for the Kroupa case of experiment 3 (chance superpositions
of binaries) shown as Hess diagrams. The $n=0$ case has been smoothed in order to allow it to show some extent perpendicular to the 1 million year isochrone.
The thick black line shows the position of the 1 million year isochrone and the thin lines the evolutionary tracks between 0 and 2 million years for the initial masses
of 25, 40, 60, 85, and 120 \Ms/. The function scaling is logarithmic and the normalization is arbitrary.
See the electronic version of the journal for a color version of this figure.}
\label{uvv_chsupbin2}
\end{figure}	

\begin{figure}
\centerline{\includegraphics*[angle=90, width=\linewidth]{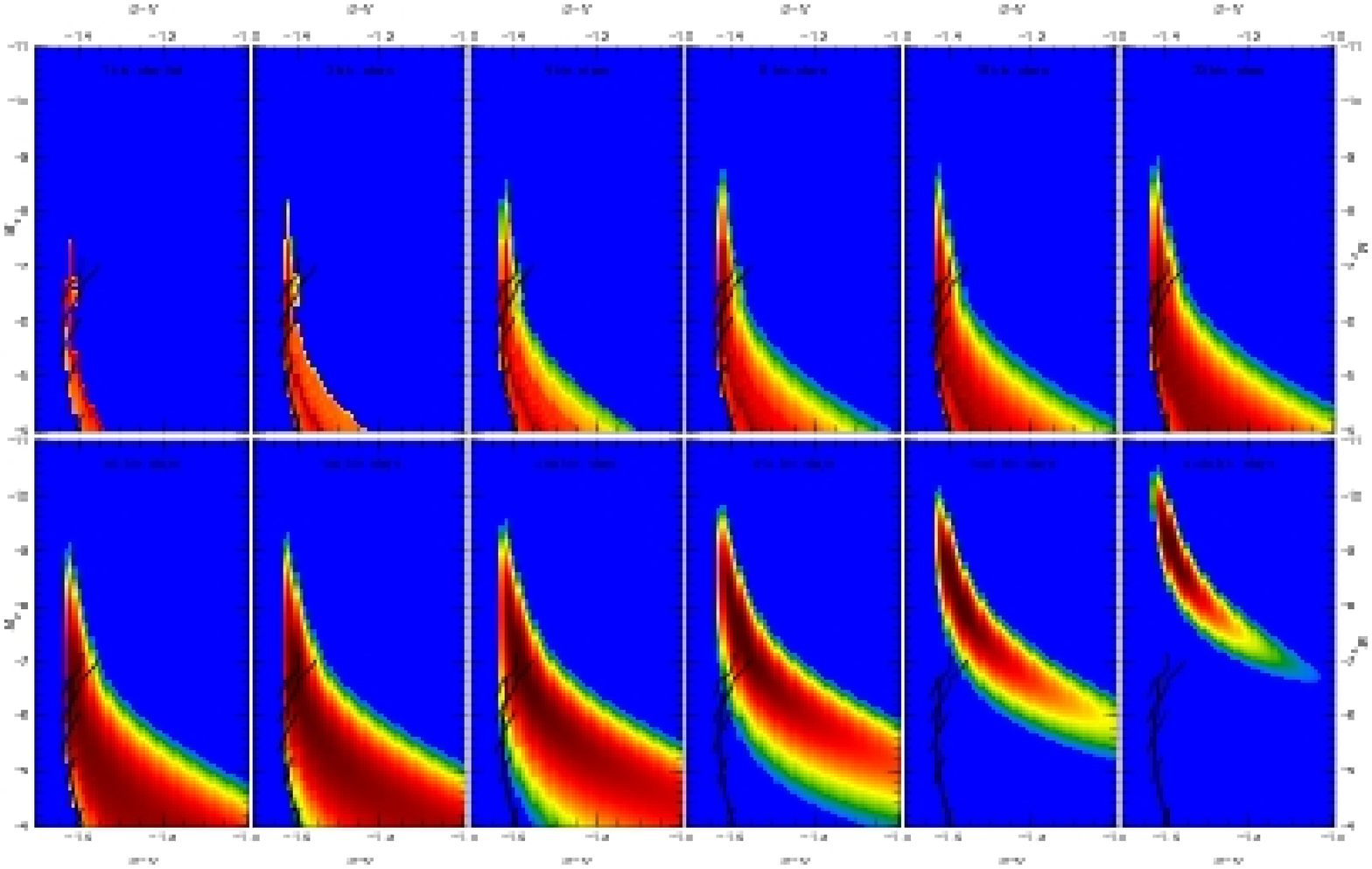}}
\caption{Top left corner of the color-magnitude density functions $g_{2^n{\rm b}}(U-V,V)$ with $n=0,11$ for the top-heavy case of experiment 3 (chance superpositions
of binaries) shown as Hess diagrams. The $n=0$ case has been smoothed in order to allow it to show some extent perpendicular to the 1 million year isochrone.
The thick black line shows the position of the 1 million year isochrone and the thin lines the evolutionary tracks between 0 and 2 million years for the initial masses
of 25, 40, 60, 85, and 120 \Ms/. The function scaling is logarithmic and the normalization is arbitrary.
See the electronic version of the journal for a color version of this figure.}
\label{uvv_chsupbin3}
\end{figure}	

	As I did for the second experiment, I show six of the resulting full color-magnitude density functions, $g_{2^n{\rm b}}(U-V,V)$ in Fig.~\ref{uvv_chsupbin1} for 
$n=1,4,7$ and the two cases (Kroupa and top heavy). The top left corner (where massive stars are located) of all the functions is shown in Figs.~\ref{uvv_chsupbin2} (Kroupa) 
and \ref{uvv_chsupbin3} (top heavy). Those three figures are quite similar to the equivalent ones for the second experiment. The most prominent differences are: 
[a] the non-zero regions of $g_{2^n{\rm b}}(U-V,V)$ are more extended in their short direction (in the diagrams, from the lower left to the upper right) compared to the 
non-zero regions of $g_{2^n{\rm s}}(U-V,V)$ due to the additional widening introduced by binaries; and [b] the extension of $g_{2^n{\rm b}}(U-V,V)$ in its long direction (from
the lower right to the upper left) is similar to that of $g_{2^n{\rm s}}(U-V,V)$ but it is slightly displaced towards lower magnitudes (higher luminosities) due to the
additional factor of two in the number of superimposed objects. The latter effect is especially noticeable for large values of $n$ (compare the lower right panels of 
Figs.~\ref{uvv_chsup2}~and~\ref{uvv_chsupbin2} or those of Figs.~\ref{uvv_chsup3}~and~\ref{uvv_chsupbin3}). Note, however, that a comparison of the two experiments assuming
the same number of superimposed individual stars implies using $g_{2^n{\rm s}}(U-V,V)$ and $g_{2^{n-1}{\rm b}}(U-V,V)$. If the comparison is done in that way one finds that for
large $n$ the top parts of the non-zero regions of the two functions end near a similar value (with the $g_{2^{n-1}{\rm b}}(U-V,V)$ one only slightly higher in luminosity) and
that the low-luminosity tail of $g_{2^{n-1}{\rm b}}(U-V,V)$ is significantly more extended than that of $g_{2^n{\rm s}}(U-V,V)$.

\begin{figure}
\centerline{\includegraphics*[width=0.62\linewidth, bb=155 175 450 720]{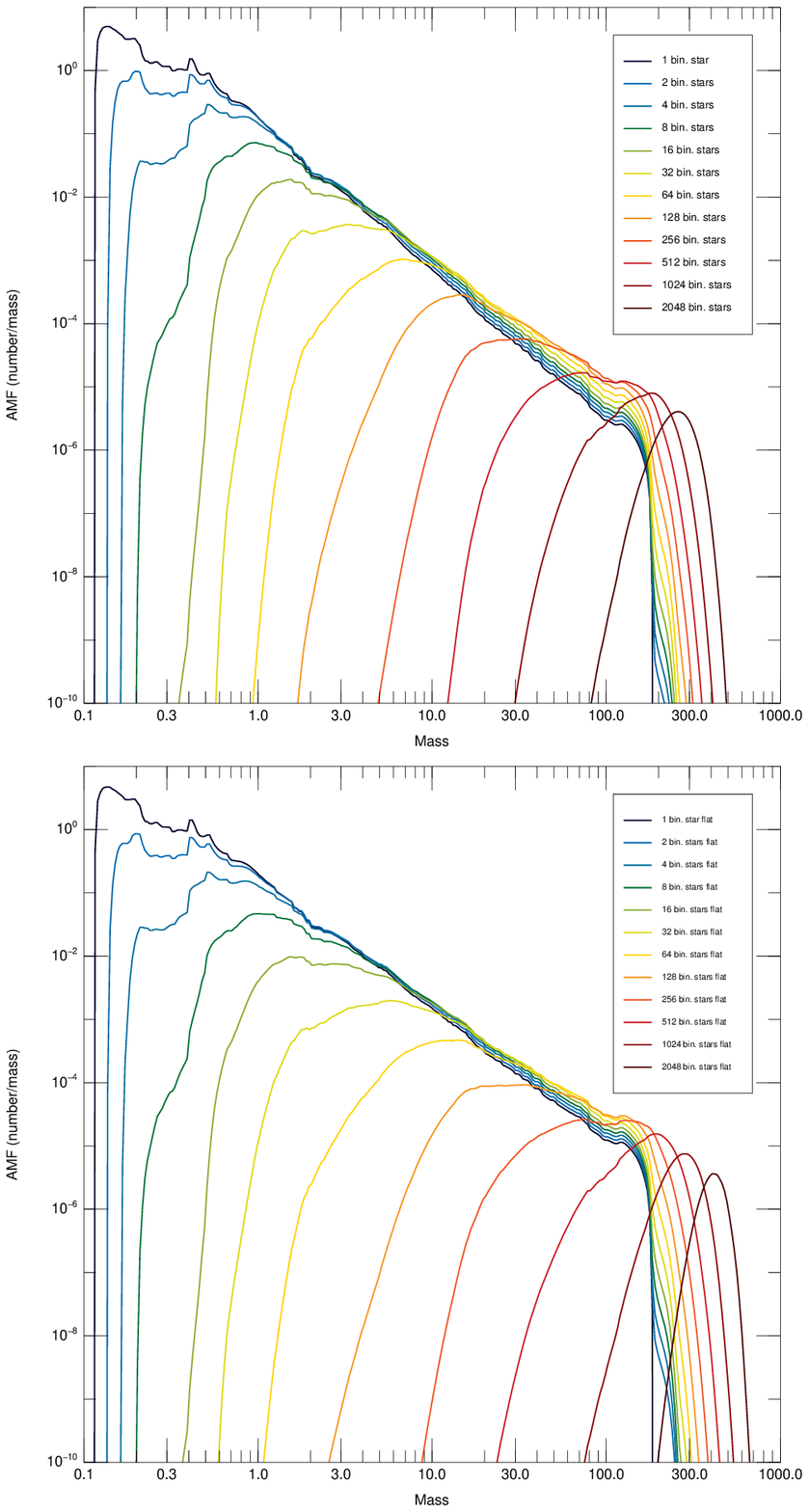}}
\caption{AMFs for experiment 3 (chance superpositions of binaries). The top panel shows the Kroupa case and the bottom panel the top-heavy case. The mass is
expressed in solar units. All AMFs are normalized to $2^{-(n+1)}$.}
\label{imf_chsupbin}
\end{figure}	

\begin{figure}
\centerline{\includegraphics*[width=\linewidth, bb=28 28 566 535]{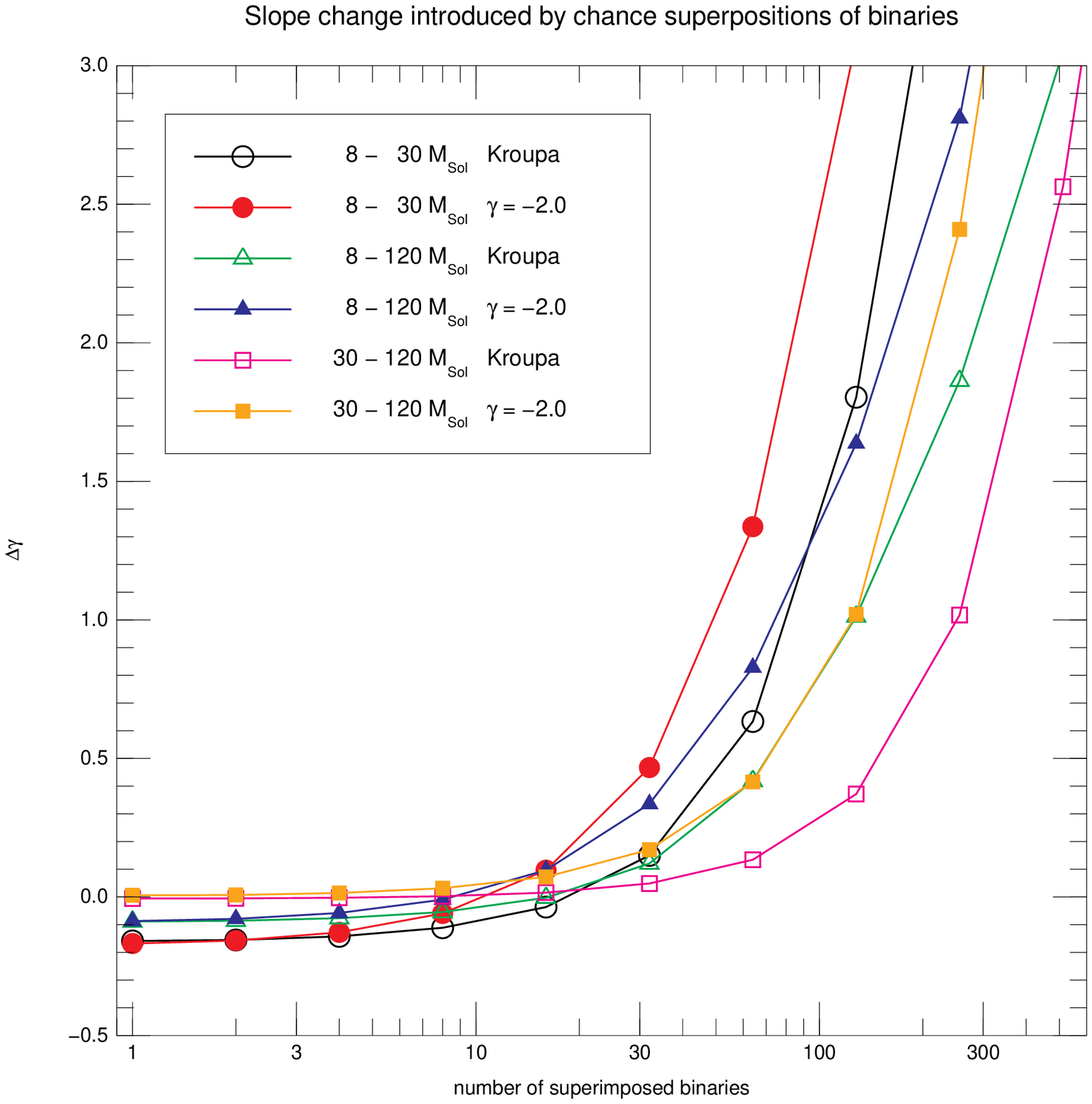}}
\caption{Slope change introduced by chance superpositions of binaries (third experiment) as a function of the number of superimposed binaries for the two 
IMF cases and for three different mass ranges.
See the electronic version of the journal for a color version of this figure.}
\label{deltaslope_chsupbin}
\end{figure}	

	The calculated AMFs for the third experiment are shown in Fig.~\ref{imf_chsupbin}, with the overall behavior in both
cases being similar to the equivalent ones for the second experiment. A more detailed comparison shows that for low values of $n$ the behavior beyond $m_{\rm max}$
is quite different: both experiments start by developing a tail from their $n=0$ case but since each initial distribution has a very different apparent upper mass limit (for 
experiment 2 it is the real $m_{\rm max}$ while for experiment 3 it is $m_{{\rm max},1{\rm b}}$) 
the fraction of stars above the limit starts from a much higher value for experiment 3 and
grows at a significantly slower pace (see Table~\ref{ratios_chsup}). For large values of $n$ the left side of $g_{2^{n-1}{\rm b}}(U-V,V)$ is quite similar to that of
$g_{2^{n-1}{\rm s}}(U-V,V)$ while its right side is similar to $g_{2^n{\rm s}}(U-V,V)$. 

	Figure~\ref{deltaslope_chsupbin} plots the change in MF slope $\Delta\gamma$ as a function of $N_{\rm sup}$ for three mass ranges. The same three regimes as in 
Fig.~\ref{deltaslope_chsup} can be distinguished. For low $N_{\rm sup}$, the effect in $\Delta\gamma$ is small and can be approximated by the value derived from the
first experiment. For intermediate values, $\Delta\gamma$ increases moderately (the mass function becomes flatter) and the effect remains correctable. Finally, for large 
$N_{\rm sup}$, $\Delta\gamma$ becomes very large and no realistic correction is possible. 

\section{Sample applications}

	In this section I apply the results of the experiments above to the IMFs derived from HST observations for two of the best studied massive young clusters in the
Local Group, NGC 3603 and R136. In order to do that, two preliminary steps are required:

\begin{enumerate}
  \item Derive the radius of an effective pixel. As previously mentioned, the value depends not only on the detector properties but also on $\Delta m$. Indeed,
	ACS/HRC on HST can separate low $\Delta m$ objects only 2 pixels apart \citep{Maizetal08a} (effective radius of 1 pixel) but for large $\Delta m$ the effective radius 
	can grow to 4-5 pixels (similar values apply to the PC of WFPC2). As a compromise, we adopt a value of 2.5 pixels.
  \item Determine the average number of stars per effective pixel, $N_{\rm epx}$, given by:

\begin{equation}
N_{\rm epx} = \left(\frac{d}{10\;{\rm pc}}\right)^2 10^{-0.4(m_{V,{\rm epx}}-M_{V,1})}.
\label{Nepx}
\end{equation}

	In Eqn.~\ref{Nepx}, $d$ is the distance to the object, $m_{V,{\rm epx}}$ is the averaged $V$-band extinction-corrected measured magnitude in a 
	single effective pixel, and $M_{V,1}$ is the absolute magnitude of a ``1-star cluster'' derived from evolutionary synthesis (calculated by deriving $M_V$ for a large 
	cluster with e.g. $N=10^6$ stars and then adding $2.5\log N$). If needed, $V$-based values can be substituted by those from other passband.
\end{enumerate}

\subsection{NGC 3603}

	NGC 3603 is the massive young cluster with the lowest extinction in the Galaxy. \citet{Moffetal94} described it as a Galactic clone of R136 without the halo due to their
similarities in age, stellar content, and central density. Note, however, that most of the mass of 30 Doradus is located outside R136, making it significantly more massive than
NGC 3603. A study with HST spectroscopy by \citet{Drisetal95}, recently extended from the ground by \citet{Meleetal07}, indicates that NGC 3603 
contains the highest concentration of well-classified O2/3 + WNha stars in the Galaxy.

	I show in Table~\ref{NGC3603data} the literature data for the distance and extinction to the core of NGC 3603. The values for the extinction are quite homogeneous and here 
we adopt $A_V = 4.50$. Also, at the location of the central cluster there is little point-to-point variation in the extinction \citep{Moff83,Melnetal89,Meleetal07},
so the assumption of a constant foreground screen appears reasonable\footnote{Strong extinction variations exist at larger radii but those do not concern us 
here.}. The disagreement on the distance to NGC 3603 is larger. Different methods and observations yield values between $\approx$~6~kpc and $\approx$~8~kpc (see \citealt{Meleetal07}
for a discussion). In order to consider
the different possibilities, I adopt three possible distances of 6, 7, and 8 kpc, which I refer to as the short, intermediate, and long distances, respectively. 
Regarding the age of the core of NGC 3603, it was previously thought to be $\approx$~3~million years based on the presence of WR stars but now it is believed that those objects
are still in their core hydrogen-burning stage (i.e. they are WNha stars, see e.g. \citealt{Crow07}) and the currently favored age is 1 million years 
\citep{Stoletal04,SungBess04}, which I adopt.

\begin{deluxetable}{lll}
\tablecaption{Distances and extinctions for NGC 3603. The uncertainties shown are standard deviations.}
\tablewidth{0pt}
\tablehead{Source & $d$ (kpc) & $A_V$ (mag)}
\label{NGC3603data}
\startdata
\citet{Moff83}     & 7.0 $\pm$ 0.5 & 4.46 $\pm$ 0.28 \\              
\citet{Melnetal89} & 7.2           & 4.47 $\pm$ 0.40 \\              
\citet{Eiseetal98} &               & 4.60            \\
\citet{dePretal99} & 6.1 $\pm$ 0.6 &                 \\
\citet{Nurnetal02} & 7.7 $\pm$ 0.2 &                 \\
\citet{Stoletal04} & 6.0 $\pm$ 0.3 & 4.50 $\pm$ 0.60 \\ 
\citet{SungBess04} & 6.9 $\pm$ 0.6 & 4.44            \\
\citet{Meleetal07} & 7.6           & 4.66            \\
\enddata
\end{deluxetable}

\begin{figure}
\centerline{\includegraphics*[width=\linewidth, bb=24 24 570 498]{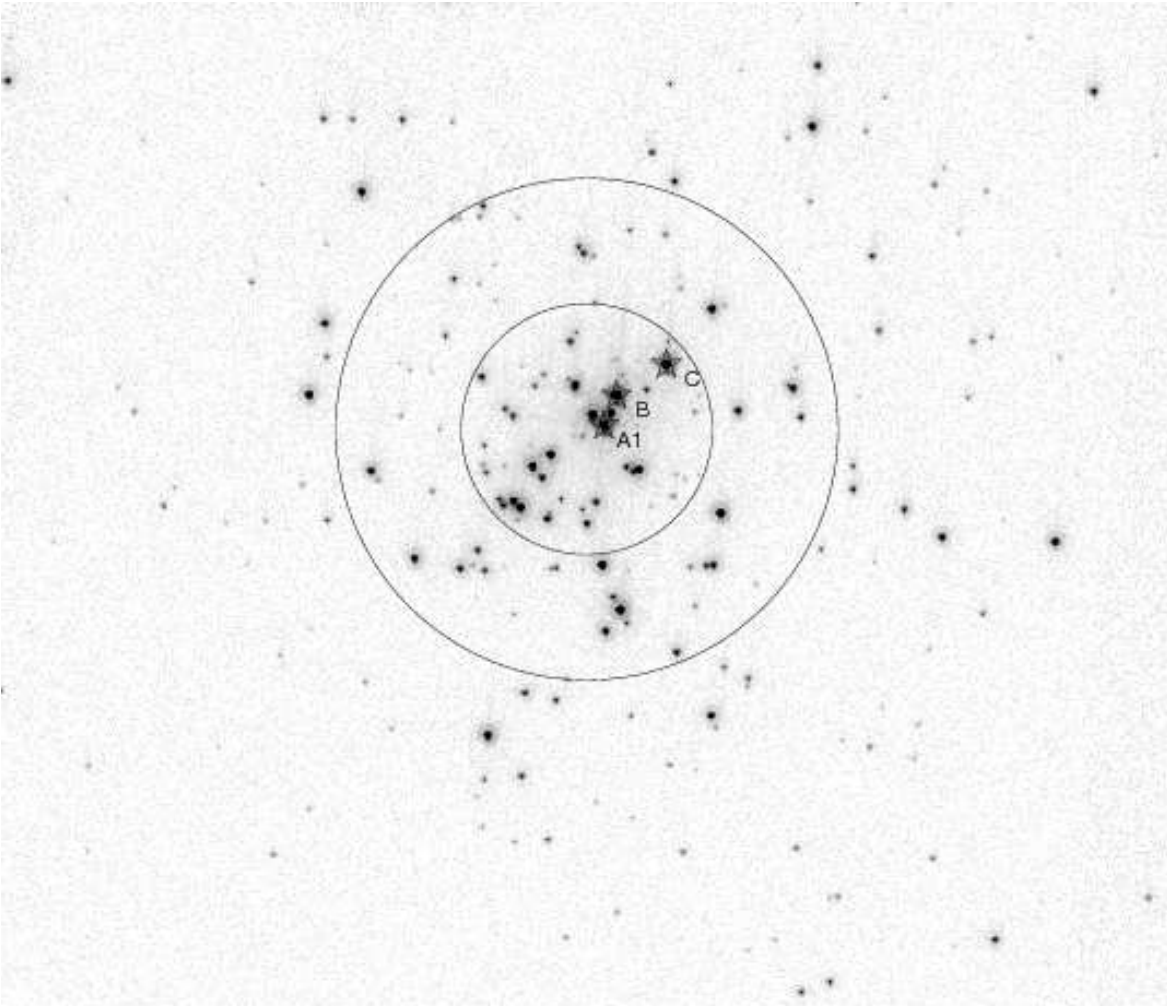}}
\caption{ACS/HRC F550M DRZ frame of NGC 3603. The three WNha objects are marked and circles with radii of 3\arcsec\ and 6\arcsec\ centered on the cluster have been drawn.
A square-root scale between 0 and 800 counts has been used in order to show both the bright and the dim stars. The pixel size is 0\farcs025, the field size is 
29\farcs3$\times$25\farcs4 (1172 px $\times$ 1016 px), and the vertical direction is 124\degr\ East of North.}
\label{ngc3603_f550m}
\end{figure}	

	NGC 3603 was observed with HST/ACS on 29 December 2005 under GO program 10602 using both the WFC and HRC detectors and several filters. Here I use the HRC F550M
data, combined with the literature information described above for the distance, extinction, and age, to study the blending effect created by real multiple systems and chance 
alignments on the NGC 3603 IMF. In a future paper we will use all the data to do a more thorough analysis that will include independent measurements of the distance and 
extinction. Four F550M exposures of 2 s each were obtained with a dithering pattern that covers the gap produced 
by the HRC occulting finger. The exposure time was chosen in order to avoid saturation of the brightest stars. MULTIDRIZZLE was used to generate a common 
geometrically-undistorted cosmic-ray-corrected drizzled (DRZ) frame, see Fig.~\ref{ngc3603_f550m}, as well as to clean the individual geometrically-distorted flat-fielded (FLT)
frames from cosmic rays and hot pixels. 

	JMAPHOT, a crowded-field photometry package specifically designed for HST images was used for the data reduction. 
First, a search is performed in the DRZ frame to find all the sources
above a certain S/N threshold. JMAPHOT finds the sources in order from bright to dim and, after each one is found, a model (the drizzled PSF for point sources, simple kernels for 
extended sources and possible leftover cosmic rays and hot pixels) is subtracted from the data and a correction applied to the weight map before the next source is searched for. 
This technique allows for the exclusion of false positives caused by complex PSFs with secondary peaks and for the detection of some of the otherwise false negatives close to 
bright sources. Once a source list is built, the coordinates are transformed back to each of the distorted, cleaned FLT frames and PSF photometry is performed on each one of 
them. Positions and fluxes of each star are fitted simultaneously; also, stars with close neighbors are grouped together and fitted at the same time. Several passes are 
performed to eliminate the small contributions from distant stars and to improve the initial PSF model (derived from \citealt{AndeKing04} with wings from Tiny Tim, 
\citealt{Kris95}) by calculating the PSF residuals. Aperture and charge-transfer efficiency corrections are applied to the flux from each individual FLT frame and the four values 
are finally combined to derive the observed VEGAMAG magnitude for each object.

	Several steps are needed to convert the observed magnitudes into masses. First, F550M values are transformed into the Johnson $V$ band by adding 0.120 mag, as
calculated for the appropriate input SED and extinction using the synthetic photometry package in CHORIZOS\footnote{The value depends on the exact SED and $A_V$ but the added 
uncertainty is of the order of only a few thousandths of a magnitude, which is good enough for our purposes.}. Second, $5\log d -5$ is subtracted to obtain the absolute 
magnitudes $V$ for each of the short, intermediate, and long distances. Finally, I use $m(V)$ to obtain the corresponding apparent masses. The first three rows in
Table~\ref{ngc3603_results} give the values I obtain for the three brightest objects in NGC 3603: A1, B, and C.

\begin{figure}
\centerline{\includegraphics*[width=0.395\linewidth, bb=210 175 400 715]{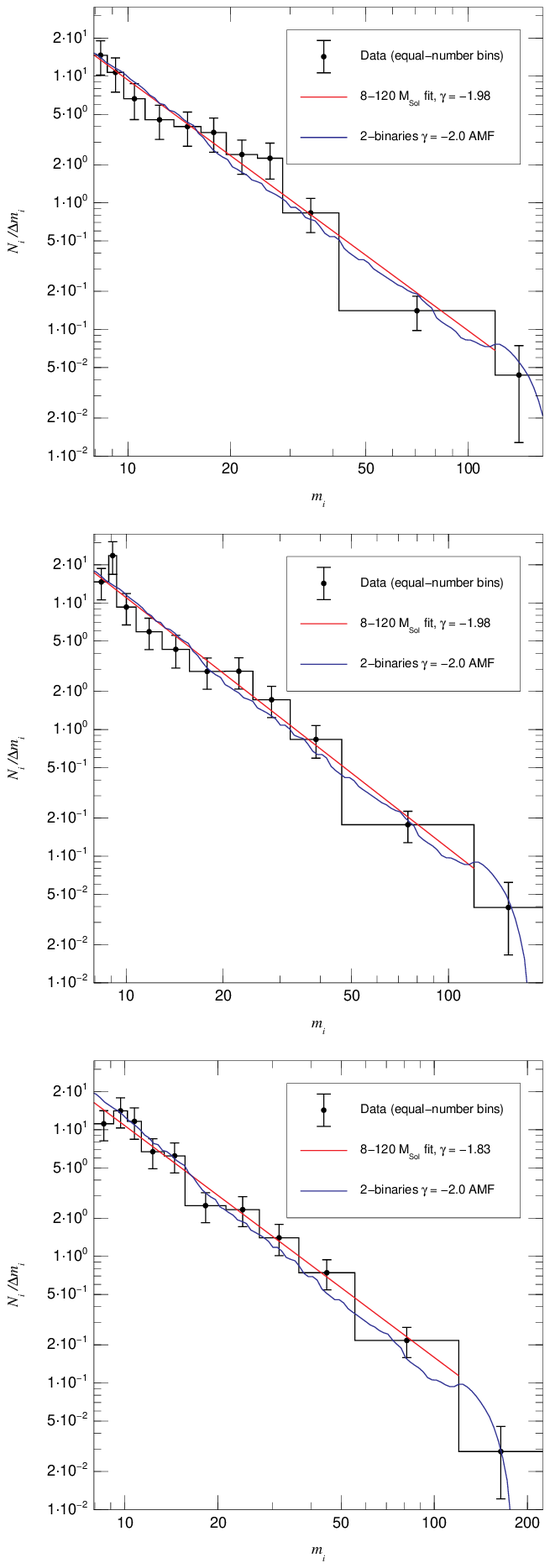}}
\caption{AMFs for NGC 3603 assuming a distance of 6.0 kpc (top), 7.0 kpc (middle), and 8.0 kpc (bottom). In each case the data, the power-law fit in the 8-120 \Ms/ range, and 
the top-heavy 2-binaries AMF are shown. Masses are in \Ms/. See the electronic version of the journal for a color version of this figure.}
\label{ngc3603_imf}
\end{figure}	

\begin{deluxetable}{lccc}
\tablecaption{Results for NGC 3603 for the three assumed distances.}
\tablewidth{0pt}
\tablehead{Quantity           & $d$ = 6 kpc        & $d$ = 7 kpc        & $d$ = 8 kpc        }
\label{ngc3603_results}
\startdata
$M_{\rm A1}$ (\Ms/)           & 159                & 189                & 217                \\
$M_{\rm B}$ (\Ms/)            & 147                & 176                & 203                \\
$M_{\rm C}$ (\Ms/)            & 105                & 130                & 152                \\
$\gamma_{\rm equal\; number}$ & $-$1.98 $\pm$ 0.16 & $-$1.98 $\pm$ 0.14 & $-$1.83 $\pm$ 0.12 \\
$\gamma_{\rm equal\; width}$  & $-$2.33 $\pm$ 0.49 & $-$2.68 $\pm$ 0.42 & $-$2.37 $\pm$ 0.39 \\
$\gamma_{\rm IMF}$            & $-$1.90 $\pm$ 0.16 & $-$1.90 $\pm$ 0.14 & $-$1.74 $\pm$ 0.12 \\
$N_{\rm epx}$ (3\arcsec)      & 0.64               & 0.88               & 1.15               \\
$N_{\rm epx}$ (6\arcsec)      & 0.21               & 0.29               & 0.38               \\
$N_{>120 M_\odot,{\rm 2s}}$   & 0.01               & 0.01               & 0.01               \\
$N_{>120 M_\odot,{\rm 1b}}$   & 2.40               & 2.83               & 3.07               \\
\enddata
\end{deluxetable}

	In order to derive the AMF I first eliminate those objects with observed masses below 8 \Ms/. Using such a large cutoff allows us not to worry about incompleteness
due to low S/N, since an 8 \Ms/ isolated star has a S/N of $\approx 35$ in a single FLT frame and of $\approx 70$ in the combined DRZ frame\footnote{Note that most
incompleteness analyses deal with two different types of non-detections, those due to low S/N of the object and those due to the proximity of bright sources. 
Since the latter is already taken
care of by our experiments, we only need to worry about the former.}.  Also, since I want to compare the results with the experiments in the previous section, those stars with
apparent masses above 120 \Ms/ (A1, B, and, for the two longest distances, C) are also excluded i.e. only the 8-120 \Ms/ range is used for the power-law fit. 
No contamination correction is applied given the small size of the field, the high 
density of stars in the cluster, and that most such objects should be dimmer than the equivalent to the 8 \Ms/ cutoff.
Finally, the AMF is fitted using bins with an equal number of objects, as
prescribed in Paper I. Different numbers of bins were tried but the slopes were left essentially unchanged, as expected from the results in that paper. The results for 10 bins are
shown in Table~\ref{ngc3603_results} and plotted in Fig.~\ref{ngc3603_imf}. For comparison purposes, a fit with 10 bins of equal width was also attempted and, as expected, the
obtained slopes were significantly different due to the bias inherent in such a method. Also, the measured uncertainties using bins of equal width were a factor of three larger (see
Table~\ref{ngc3603_results}). Note that for the three distances $\gamma_{\rm equal\; number}$ is compatible (less than 1 sigma in two cases, less than 1.5 sigmas in the other) 
with the value we have used for our top-heavy experiments ($\gamma = -2.0$)\footnote{Of course, the observed $\gamma$ is that of the AMF, not the IMF, but, as we will see below, both
should be very similar in this case.}, so that will be the assumed slope for the rest of this subsection.

	Finally, to correct for the effects of unresolved binaries and/or chance superpositions, we need to calculate $N_{\rm epx}$ using Eqn.~\ref{Nepx}. With the
evolutionary synthesis module of CHORIZOS I obtain that the absolute $V$ magnitude of a solar-metallicity 1-million-year old top-heavy ``1-star cluster'' is $-0.75$. The other 
quantity needed is the observed average magnitude in an effective pixel, which will depend on the aperture radius chosen. The DRZ field has a size of 29\farcs3$\times$25\farcs4 but 
69\% of the flux is contained within a radius of 3\arcsec\ and 94\% within a radius of 6\arcsec. Therefore, the area enclosed within those two radii provide reasonable descriptions of 
the center of NGC 3603 and of the region that contains most of the light from the core, respectively. The resulting $N_{\rm epx}$ are shown in Table~\ref{ngc3603_results}. Depending 
on the distance and radius selected, I find that, on average, each point source in NGC 3603 contains at most 1.15 individual stars due to chance superpositions with the 
assumption that all systems are single and not unresolved binaries. Therefore, chance superpositions are not an important issue for the NGC 3603 HRC data and, indeed, most effective 
pixels should be empty. Of course, if most point sources are binaries the results of the first experiment are still relevant.

	What do our experiments tell us about the corrections required to analyze the AMF in terms of the real IMF of NGC 3603? With the observed values of $N_{\rm epx}$, 
$f_{a,1{\rm b}}$ (1 binary) should be the closest approximation to reality if indeed most stars are binaries (note that for binaries the values of $N_{\rm epx}$ have to be reduced by
a factor of two and that those values represent the average number of binaries per effective pixel). As a comparison, I also discuss the case of $f_{a,2{\rm s}}$ (2 single stars),
which is less realistic but is the closest choice among the results of the second experiment. Regarding $\Delta\gamma$, both AMFs indicate that the measured slope 
for the range 8-120 \Ms/ will be slightly steeper than the real IMF ($\Delta\gamma = -0.086$ $f_{a,2{\rm s}}$ and by $\Delta\gamma = -0.087$ for $f_{a,1{\rm b}}$). Applying the
correction $-\Delta\gamma$ we arrive at the final values for the IMF slope, $\gamma_{\rm IMF}$, shown in Table~\ref{ngc3603_results}. For the short and intermediate distances, I obtain
a slope close to the $-2.0$ while for the long distance the slope is slightly steeper. The values are in good agreement with the recent result by \citet{Stoletal06}, who measured an IMF 
for NGC 3603 with $\gamma$ = -1.89 $\pm$ 0.14 between 7\arcsec\ and 20\arcsec\ (note, however, that their $m_{\rm lower}$ is significantly lower than 8 \Ms/). 

	The result is different for the expected number of ultramassive objects, $N_{>120 M_\odot}$, shown in Table~\ref{ngc3603_results}. For the single-star 
case, the expected number is almost zero. For the binary case, the predicted numbers are $\sim$2 for the short distance and $\sim$3 for the intermediate and long distances.
In other words, {\it if the real stellar upper mass limit is 120 \Ms/, we would expect two or three stars with
apparent masses beyond that limit if most objects are unresolved binaries} and that is precisely the number that is observed. 

	What else do we know about A1, B, and C, the three brightest point sources in NGC 3603? \citet{Drisetal95} identified them as the source of the Wolf-Rayet part of the
composite WR+O integrated spectrum of the core, as had been already partially suggested by \citet{Walb73b}. Since hydrogen is present and there is intrinsic absorption in the high-order
Balmer lines, they are
classified as WN6ha, implying that they are the more luminous and massive cousins of the O3 stars in their immediate vicinity. Therefore, from the point of view of their spectral 
classification they are good candidates to have masses above 120 \Ms/ if such stars really exist. However, it is also possible that they are unresolved multiple systems composed of
individual stars with masses in the range 70-120 \Ms/ which, from what we currently know, should have the same or similar spectral type (see e.g. \citealt{Bonaetal04} and note
that their values are present-day masses, not initial ones). 

	One way to decide which of the two options is correct is to search for radial velocity variations and possible eclipses, which is what \citet{Moffetal04} did. They found that 
A1 is an eclipsing binary with a 3.7724 day period. Olivier Schnurr and Tony Moffat (private communication) have recently measured the radial velocity variations of A1 and obtained masses 
of 114 $\pm$ 30 \Ms/ and 84 $\pm$ 15 \Ms/, respectively. Interestingly, if we take those masses to be the initial ones, a 114 \Ms/ + 84 \Ms/ unresolved object should have an apparent mass 
of 157 \Ms/; assuming some stellar mass loss during their lifetimes may add up to a few tens of solar masses to that value. The observed evolutionary masses lie between 159 \Ms/ and 
217 \Ms/, which is in very good agreement with the Schnurr and Moffat result.

	Schnurr and Moffat (in preparation) have also found that C is a binary with an 8.92 day period, so at least two of the WN objects in NGC 3603 are not single. 
B shows no signs of radial velocity variations so far but that does not mean that it is not a binary,
since most massive binaries have separations (and some, also inclinations) that make their motions hard to detect. On the other hand, both A1 and C are overluminous in X-rays
\citep{Moffetal02}, usually a sign of binarity since intense X-ray emission is produced in wind-wind interactions but B is normal. Then again, that does not preclude B from being a 
binary with a separation large enough not to emit extra X-rays but still small enough not to have its two components resolved by ACS/HRC.

	In summary, our analysis of NGC 3603 has found out that the existence of 2-3 objects with apparent masses above 120 \Ms/ would be expected as the result of the observed 
number of stars in the cluster and its IMF below that limit without having to invoke the existence of stars with real masses above 120 \Ms/. The underlying hypothesis is that most or 
all massive stars in NGC 3603 are members of binary systems. Independent data have shown that indeed two of those stars are binaries and the currently available results for the 
third one leave room for the decision to go either way. The measurement of $\gamma$ could be revised slightly when the additional ACS data are processed. When we do that in a future
paper, we will apply CHORIZOS to calculate individual extinction corrections for each star, derive a distance from the data, and do a comparison between the resulting color-magnitude
diagram with the one in Fig.~\ref{uvv_bin}.

\subsection{R136}

	From a historical point of view, R136 (HD 38268), the core of 30 Doradus, is an excellent case study on the problems caused by multiplicity, either real or induced by chance
superpositions. \citet{Walb73b}, noting the WR features in its spectrum, compared it to NGC 3603 and suggested that it could be a Trapezium-like system made out of Wolf-Rayet and
other early-type O stars. On the other hand, \citet{Cassetal81} dismissed that the central object, R136a, could be made out of ``30 O3 or WN3 within a space of 0.5 arcsec (0.1 pc)''
and proposed instead that it was a single supermassive star with 2500 \Ms/. A few years later that alternative was shown to be false first by \citet{MoffSegg83} by detecting spectral 
variations within R136 and later by \citet{WeigBaie85} by using speckle interferometry to resolve R136a into eight individual sources, something which was later confirmed with HST 
\citep{Weigetal91}. 

	Several later HST studies have analyzed the stellar composition of R136, of which the most complete is that of \citet{MassHunt98}. They obtained FOS spectroscopy of
all the bright point sources in the core and combined it with PC (WFPC2) photometry to derive a color-magnitude diagram of the massive stars. In a follow-up paper, 
\citet{Massetal02} observed four spectroscopic eclipsing binaries and measured their masses. They also detected another five eclipsing binaries, indicating that massive binaries are 
also common in R136. From their data, \citet{MassHunt98} measured a Salpeter- or Kroupa-like IMF with $\gamma$ between $-2.3$ and $-2.4$ for the range 15-120 \Ms/. That value, however,
does not correspond to R136 alone but also includes the inner part of the 30 Doradus halo. In any case, for lack of better information, it will be the value assumed here.

\begin{figure}
\centerline{\fbox{\includegraphics*[width=\linewidth]{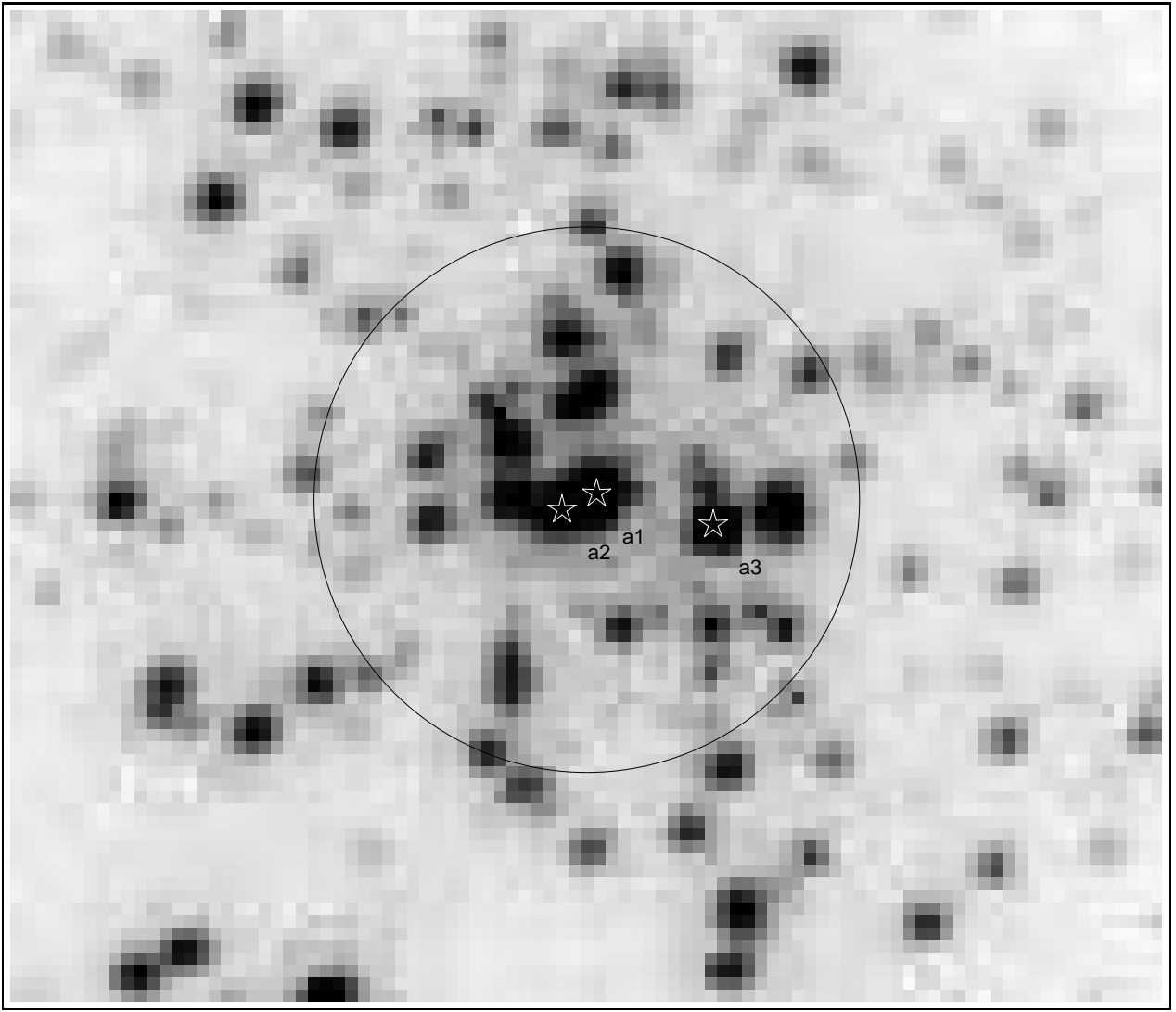}}}
\caption{PC (WFPC2) F555W image of R136. The three WNha objects are marked and a circle with radius of 1\arcsec\ centered on the cluster has been drawn.
A square-root scale between 0 and 400 counts has been used in order to show both the bright and the dim stars. The pixel size is 0\farcs0455, the field size is 
4\farcs23$\times$3\farcs64 (93 px $\times$ 80 px), and the vertical direction is 37.8\degr\ West of North. The physical size (1.0 pc $\times$ 0.9 pc) is similar to that of
Fig.~\ref{ngc3603_f550m}.}
\label{r136_f555w}
\end{figure}	

	R136 is roughly seven times farther away than NGC 3603. Also, 30 Doradus has never been observed with ACS (either WFC or HRC) and the best data currently available were
obtained with the PC on WFPC2 (that was what \citealt{MassHunt98} used), which has a factor of 1.6 lower pixel size than HRC. Therefore, if R136 and NGC 3603 had similar stellar
densities at their centers, one would expect crowding to be $(1.6\times 7)^2 \approx 125$ times worse in the former. The difference in resolution is readily apparent when one 
compares Figs.~\ref{ngc3603_f550m}~and~\ref{r136_f555w}.

	Let us quantify how important crowding is within a 1\arcsec-radius circle centered on R136. Doing aperture photometry on the F555W data, obtaining $A_V$ from the data in
\citet{MassHunt98}, and using CHORIZOS to estimate the conversion to the $V$ passband, 
I derive an extinction-corrected $m_V$ of 9.29. Assuming an effective pixel radius of 2.5 real pixels I obtain 
$m_{V,{\rm epx}}$ = 14.01. For assumed values of 1 million years (age)\footnote{R136 could hardly be significantly younger than that because it has already cleared of dense gas a large 
region around it. Also, if its age were older than 2 million years, its most massive stars would no longer be of WNha type.}, LMC (metallicity), and Kroupa (IMF), the value 
$M_{V,1} = 0.79$ is obtained. Finally, assuming a distance of 50~kpc, Eqn.~\ref{Nepx} yields $N_{\rm epx} = 129$. The value is relatively robust to small variations in the assumed 
parameters. Therefore, the $n=7$ results of the second experiment (if binaries are not 
abundant in R136) and the $n=6$ results of the third experiment (if binaries dominate) should be the reference for the analysis of the PC data of R136. This means that all the 
objects within the circle in Fig~\ref{r136_f555w} have to be the superposition of a large number of stars (the majority of them being of low mass).

	Using $f_{a,64{\rm b}}$ we obtain that 1.6\% of the 77 effective pixels within 1\arcsec\ are expected to contain one object above 120 \Ms/. The equivalent value for
$f_{a,128{\rm s}}$ is 0.6\%. This translates into 1.3 and 0.5 stars, respectively\footnote{Note that the expected number of ultramassive stars is lower for R136 than for NGC 3603,
despite the facts that the regions considered here have the same extinction-corrected absolute magnitude and that R136 is more crowded. The explanation resides in the different IMFs 
used for the calculation, Kroupa for R136 and top-heavy for NGC 3603.}. 
How does this compare with the observations? Using the currently preferred low-temperature-scale 
values of \citet{MassHunt98}, the three most massive stars in the region are R136a1, a2, and a3, with masses of 136 \Ms/, 122 \Ms/, and 120 \Ms/, respectively. Those observed masses
agree very well with the values predicted by the experiments in this paper, especially if binaries dominate. It is also interesting to point out that three of the four most massive
objects with spectral types in the larger region explored by \citet{MassHunt98} 
are within the 1\arcsec\ radius considered here. One explanation would be that the most massive stars form
preferentially in very dense environments. However, there are five other stars with similar spectral types (WNha and O3 If*/WN) outside the most crowded regions with masses only
slightly lower (between 103 \Ms/ and 120 \Ms/). Hence, a reasonable alternative is that all of those stars have similar masses and that the extra measured mass in R136a1, a2, and a3
is simply due to the increased likelihood of chance superpositions there.

	R136 was the cluster used by \citet{WeidKrou04} and \citet{Koen06} and one of the clusters in the sample used by \citet{OeyClar05} to derive an $m_{\rm max}$
around 150 \Ms/ (with varying degree of uncertainty around the value). All of those works used the \citet{MassHunt98} values for the masses which, even for the low-temperature scale, 
include at least a star with a mass of 136 \Ms/ (R136a1). Of course, the existence of a single star with such a mass invalidates the possibility that $m_{\rm max}$ is lower than
that value. However, if, as we have seen, the certain existence of chance superpositions and the likelihood that R136a1 is a binary\footnote{Note that R136a1+a2 is
not likely to be a bound system, as they are separated by at least 15\,000 AU.} (as most massive stars are) are considered, then the real mass of the heaviest component in 
R136a1 could be 120 \Ms/ or even less.

	What about deriving the true IMF slope from the observed AMF at the center of R136? Figs.~\ref{deltaslope_chsup}~and~\ref{deltaslope_chsupbin} indicate that 
$\Delta\gamma$ is in the range between 0.8 and 1.5 for 8-120 \Ms/ in the region within 1\arcsec. That correction is too large and too dependent on the assumptions to be applied with 
confidence, so with the current data it is not possible to know the real IMF slope at the very center of R136. However, note that if a factor of 2 in (linear) effective pixel size is 
achieved, a dramatic change takes place. The relevant functions then are $f_{a,32{\rm s}}$ and $f_{b,16{\rm s}}$ and $\Delta\gamma$ becomes $\approx 0.10$, which is a believable
correction. Note also that in such a case, with $\approx 300$ effective pixels and with a S/N large enough to have 100-150 of them populated with detected stars, the number of
observed objects will be enough to derive a bias-free meaningful IMF ($\sigma_\gamma \le$ 0.2) if the techniques in Paper I are used (see section 5 in \citealt{MaizUbed05}).

	Our summary for R136 is similar to that for NGC 3603. Even though there are stars with apparent masses above 120 \Ms/, their detection can be explained by unresolved stars 
with real masses below that value. The main differences are that here the measured values are only slightly above the limit and that here the unresolved stars are a combination of
chance superpositions and (likely) real multiple systems. $\gamma$ cannot be reliably measured with the current data for the innermost regions of R136 but a moderate improvement in
the quality (within the capabilities of HRC/ACS) should be sufficient to do the job.

\section{Discussion}

	I have presented in this paper a simple method to quantify the effect of unresolved multiple systems and chance alignments on the heavy part of the mass functions of young stellar 
clusters and applied it to NGC 3603 and R136. Unresolved multiple systems introduce an intrinsic spread in the observed color-magnitude diagrams and can produce a population
of AUMSs in a rich cluster. On the other hand, their effect on the apparent massive star MF slope is relatively small. Chance alignments can also produce AUMSs
but only for large $N_{\rm sup}$. As opposed to unresolved multiple systems, their effect on the AMF slope can be large.

	To my knowledge, no similar method had been previously developed for those same circumstances. For the effect of binaries in the IMF of
low-mass stars in the field and in globular clusters, see \citet{Krouetal91} and \citet{Solletal07}, respectively. For a similar (but more restricted) analysis to the one in this
paper applied to intermediate-age clusters, see \citet{SagaRich91}. Previous analyses for young clusters sometimes include a simplified correction with equal-mass binaries (e.g.
\citealt{Stoletal06}) but in most cases binaries are included in the incompleteness analysis. An incompleteness test adds a large number of fake stars to the observed field and runs
the output through the star detection algorithm to see how many of the fake objects are recovered. The percentage of unrecovered objects is then used to calculate an incompleteness
correction. Such an analysis in reality is measuring two different effects: the weakest objects (those close to the detection limit) may be missed, even in an uncrowded
region, either because Poisson/background fluctuations or read noise place them below the detection limit or because the details of the detection algorithm make it stop before
reaching it. Alternatively, some brighter objects are missed because they are located close to an even brighter star that hides them under its PSF. The first effect (low S/N
objects) should be quantifiable by studying the detector properties and the expected S/N as a function of magnitude. The second effect is more complex because it depends on the PSF
(which may be or may be not well characterized) and, more importantly, the stellar distribution. For that reason, incompleteness tests need to be run on a specific dataset.
Furthermore, a classical incompleteness test does not address the issue that the added fake stars may significantly modify the measured magnitude of the detected, brighter star, and,
thus, alter the results of the mass function. The technique presented here does not attempt to substitute the correction of the first effect (low S/N objects) provided by
incompleteness tests but only the second. It does so by [a] correcting the measured slope after it is fitted to the observed mass function rather than modifying the observed mass 
function before fitting it and [b] including the effect of the different mass detected when two stars are superimposed and not only counting the fraction of undetected objects. 
It also has the advantage of being easily quantifiable and applicable to the planning of future observations.

	Having analyzed in detail NGC 3603 and R136, it is interesting to point out that no end-products of runaway collisions are seen there despite those two clusters being
mentioned as likely candidates for them by \citet{Portetal04a}. 
There are several possible explanations: [a] Runaway collisions do take place but the remaining products evolve so fast that 
those in NGC 3603 and R136 have already become black holes (this would require lifetimes of $\approx$ 1 million years). [b] Core collapse happens at a later stage,
so the required time and density conditions are not satisfied in these clusters and ultramassive objects form only as a result of the merger of evolved objects. [c] Core collapse
never happens and ultramassive objects 
never form. [d] The output of a massive stellar merger that would lead to an ultramassive stars is so unstable that it loses its excess mass almost immediately. Regarding the 
latter possibility, it is notorious that very massive stars are very difficult to model and that our current knowledge of them is limited. Nevertheless, we know that there could be 
different mechanisms at work that would impede the formation of an ultramassive star via a merger: the resulting object may be directly above 
Eddington's limit \citep{GlebPols07}, so close to it that strange-mode instabilities overwhelm it \citep{Town07}, or may always rotate so fast that it surpasses the
$\Omega\Gamma$ limit \citep{MaedMeyn00}.

	This article provides further evidence for the existence of a stellar upper mass limit at solar or near-solar metallicities. Furthermore, the analyzed data indicate that 
$m_{\rm max}$ may be as low as 120 \Ms/. The situation for Population III
stars is expected to be different, likely because the reduced opacity should place Eddington's limit at higher luminosities and, hence, masses (see e.g. \citealt{ZinnYork07} and 
references therein).

\begin{acknowledgements}

Support for this work was provided by the Spanish Government Ministerio de Educaci\'on y Ciencia through grant AYA2004-08260-C03, grant AYA2007-64712,
and the Ram\'on y Cajal Fellowship program and co-financed with FEDER funds. I would like to thank Tony Moffat and Phil Massey for providing me access 
to their work before publication; Nolan Walborn, Tony Moffat, and an anonymous referee for useful comments to previous versions of the manuscript; and 
Enrique P\'erez and Miguel Cervi\~no for fruitful conversations on this topic.

\end{acknowledgements}

\bibliographystyle{apj}
\bibliography{general}

\end{document}